\documentclass[twocolumn]{aa}

\usepackage{graphicx}
\usepackage{lscape}
\usepackage{natbib}
\bibpunct{(}{)}{;}{a}{}{,} 
\newcommand\msun{\rm\,M_\odot}
\newcommand\M{\rm M}

\usepackage{txfonts}
\begin{document}

   \title{On central black holes in ultra-compact dwarf galaxies}
\author{
          S. Mieske
          \inst{1}
          \and
          M. J. Frank \inst{2}
          \and
          H. Baumgardt
          \inst{3}
          \and N. L\"{u}tzgendorf \inst{4}
          \and N. Neumayer \inst{4}
          \and M. Hilker \inst{4}}

   \offprints{S. Mieske}

       \institute{
               European Southern Observatory, Alonso de Cordova 3107, 
Vitacura, Santiago, Chile
          \and
               Landessternwarte, Zentrum f\"ur Astronomie der 
Universit\"at Heidelberg, K\"onigsstuhl 12, 69117 Heidelberg, Germany
          \and
               School of Mathematics and Physics, The University of 
Queensland, Brisbane, QLD 4072, Australia
\and European Southern Observatory, Karl-Schwarzschild-Stra{\ss}e 2, 
85748 Garching b. M\"unchen, Germany
          }

   \date{}

 
\abstract 
{The dynamical mass-to-light (M/L) ratios of massive ultra-compact
  dwarf galaxies (UCDs) are about 50\% higher than predicted by
  stellar population models.}
{Here we investigate the possibility that these apparently elevated
  M/L ratios of UCDs are caused by a central black hole (BH) that heats up the
  internal motion of stars. We focus on a sample of $\sim$50
  extragalactic UCDs from the literature for which velocity dispersions
  and structural parameters have been measured.}
{To be self-consistent in our BH mass estimates, we first redetermine
  the dynamical masses and M/L ratios of our sample UCDs, using
  up-to-date distance moduli and a consistent treatment of aperture
  and seeing effects. On average, the homogeneously redetermined
  dynamical mass and M/L ratios agree to within 5\% with previous
  literature results. We calculate the ratio $\Psi=(M/L)_{\rm
    dyn}/(M/L)_{\rm pop}$ between the dynamical and the stellar
  population M/L for an assumed age of 13 Gyr. $\Psi>1$ indicates an
  elevated dynamical M/L ratio, suggesting dark mass on top of a
  canonical stellar population of old age. For all UCDs with $\Psi>1$
  we estimate the mass of a hypothetical central black hole needed to
  reproduce the observed integrated velocity dispersion}
{ Massive UCDs ($\M>10^7\msun$) have an average $\Psi = 1.7\pm 0.2$,
  implying notable amounts of dark mass in them. We find that, on
  average, central BH masses of 10-15\% of the UCD mass can explain
  these elevated dynamical M/L ratios. The implied BH masses in UCDs
  range from several $10^5 \msun$ to several $10^7 \msun$. In the
  $M_{\rm BH}-$Luminosity plane, UCDs are offset by about two orders
  of magnitude in luminosity from the relation derived for
  galaxies. Our findings can be interpreted such that massive
    UCDs originate from progenitor galaxies with masses around $\sim
    10^9 \msun$, and that those progenitors have SMBH occupation
    fractions of $\sim$60-100\%. The suggested UCD progenitor masses
  agree with predictions from the tidal stripping scenario. Also, the
  typical BH mass fractions of nuclear clusters in such $\sim 10^9
  \msun$ galaxy bulges agree with the 10-15\% BH fraction estimated
  for UCDs. Lower-mass UCDs ($\M<10^7\msun$) exhibit a bimodal
  distribution in $\Psi$, suggestive of a coexistence of massive
  globular clusters and tidally stripped galaxies in this mass
  regime.}
{Central BHs as relict tracers of tidally stripped progenitor galaxies
  are a plausible explanation for the elevated dynamical M/L ratios of
  UCDs. Direct observational tests of this scenario are suggested. }

\titlerunning{Black holes in UCDs}


   \maketitle 
%

\section{Introduction}

Ultra-compact dwarf galaxies (UCDs) are compact stellar systems with
sizes $\lesssim 100$ pc and stellar masses between $2\times$10$^6$ and
10$^8 \msun$. They were discovered about 15 years ago (Minniti et
al.~\citeyear{Minnit98}, Hilker et al.~\citeyear{Hilker99}, Drinkwater
et al.~\citeyear{Drinkw00}) in the course of spectroscopic surveys of
the Fornax cluster. They were identified as cluster members unresolved
in ground-based imaging, but having luminosities an order of magnitude
brighter than omega Centauri (NGC 5139), up to $M_V \sim -13.5$
mag. Subsequent HST imaging revealed that their intrinsic sizes are
typically a few tens of pc (e.g. Drinkwater et
al.~\citeyear{Drinkw03}, Ha\c{s}egan et al.~\citeyear{Hasega05},
Mieske et al.~\citeyear{Mieske06}, Hilker et al.~\citeyear{Hilker07}),
in between typical star cluster and dwarf galaxy scales. In the
luminosity-size plane, UCDs occupy the region between star clusters
and dwarf galaxies (e.g. Misgeld \& Hilker~\citeyear{Misgel11}). Not
surprisingly, UCDs are thus interpreted either as tidally threshed
nuclei of dwarf galaxies (Phillipps et al.~\citeyear{Philli01}, Bekki
et al.~\citeyear{Bekki03}, Pfeffer \& Baumgardt~\citeyear{Pfeffe13},
Strader et al.~\cite{Strade13}), or as the most massive star clusters
of their host galaxy's globular cluster system (e.g. Fellhauer \&
Kroupa~\citeyear{Fellha02} \&~\citeyear{Fellha05}, Mieske et
al.~\citeyear{Mieske02}, ~\citeyear{Mieske12}).

The transition from UCDs to star clusters is smooth in the mass-size
plane (e.g. Forbes et al.~\citeyear{Forbes08}, Misgeld \&
Hilker~\citeyear{Misgel11}). A number of studies have found and
discussed that intrinsic sizes of compact stellar systems start to
scale with luminosity for masses above a few million solar masses
(e.g. C\^{o}t\'{e} et al.~\citeyear{Cote06}, Dabringhausen et
al.~\citeyear{Dabrin08}). This has been attributed to the
  existence of a maximum stellar surface density
  (e.g. Murray~\citeyear{Murray09}, Hopkins et
  al.~\citeyear{Hopkin10}), probably regulated by feedback from
  massive stars. At the same limiting mass, the average dynamical M/L
ratios of compact stellar systems start to increase to $\sim$50\%
beyond the value expected for a canonical stellar mass function
(Dabringhausen et al.~\citeyear{Dabrin08}, Ha\c{s}egan et
al.~\citeyear{Hasega05}). Because of these two changes in scaling
relations, 2$\times 10^6 \msun$ has often been adopted as a limit
between massive star clusters and UCDs (Misgeld \&
Hilker~\citeyear{Misgel11}, Dabringhausen et al.~\citeyear{Dabrin08},
Mieske et al.~\citeyear{Mieske08}, Forbes et al.~\citeyear{Forbes08}),
 although note the caveat that a one-dimensional separation in
  mass is probably too simplistic (e.g. Brodie et
  al. \citeyear{Brodie12}).

\subsection{The elevated dynamical M/L of UCDs}

The elevated dynamical M/L in the regime of UCDs have received
significant attention in the literature (Ha\c{s}egan et
al.~\citeyear{Hasega05}, Dabringhausen et
al.~\citeyear{Dabrin09},~\citeyear{Dabrin10} \& ~\citeyear{Dabrin12},
Baumgardt \& Mieske~\citeyear{Baumga08}, Mieske \&
Kroupa~\citeyear{Mieske08b}, Taylor et al.~\citeyear{Taylor10}, Frank
et al.~\citeyear{Frank11}, Strader et al.~\citeyear{Strade13}). One
possible explanation for this rise is a change of the initial stellar
mass function (IMF) in these extreme environments. An IMF variation
has been proposed in the form of an overabundance of massive stellar
remnants (Murray~\citeyear{Murray09}, Dabringhausen et
al.~\citeyear{Dabrin09},~\citeyear{Dabrin10}), or of low-mass stars
(Mieske \& Kroupa~\citeyear{Mieske08b}). The latter case would be
similar to recent suggestions of a bottom-heavy IMF in giant
ellipticals (van Dokkum \& Conroy~\citeyear{vandok10}, Conroy \& van
Dokkum~\citeyear{Conroy12}). Dabringhausen et
al.~(\citeyear{Dabrin12}) presented circumstantial evidence for a
top-heavy IMF in UCDs, while a first test for a bottom-heavy stellar
mass-function in two extremely high-M/L UCDs yielded a non-detection
(Frank et al. in preparation, cf. Frank~\citeyear{Frank12}).

Also, a cosmological origin for the elevated M/L of UCDs in the form
of dark matter has been discussed (Ha\c{s}egan et
al.~\citeyear{Hasega05}, Goerdt et al.~\citeyear{Goerdt08}, Baumgardt
\& Mieske~\citeyear{Baumga08}). In case UCDs are threshed nuclei, they
may still reside in the (remnant) dark matter haloes of the progenitor
galaxies. However, in order for dark matter to have a measurable
impact on the stellar kinematics of such compact objects the central
density would have to be orders of magnitude higher than derived for
dwarf spheroidal galaxies (e.g. Gilmore et al.~\citeyear{Gilmor07},
Dabringhausen et al.~\citeyear{Dabrin10}) or expected from theoretical
dark matter profiles when assuming realistic total halo masses
(Murray~\citeyear{Murray09}). Goerdt et al.~(\citeyear{Goerdt08}) and
Baumgardt \& Mieske~(\citeyear{Baumga08}) discussed that this problem
may be alleviated, if the central dark matter concentration was
enhanced during the formation of the progenitor nuclei via the infall
of gas. In this context it is worth noting that Frank et
al.~(\citeyear{Frank11}) did not find any evidence for an extended
dark matter halo in Fornax UCD3 based on its velocity dispersion
profile (see also next subsection).

\subsection{Black holes as a possible explanation for elevated M/L}

Another obvious possibility for additional dark mass in UCDs is the
presence of a massive central black hole. Qualitatively, a central
black hole of a given mass will enhance the global velocity dispersion
of a compact stellar system much more than the same amount of mass
distributed uniformly (Merritt et al.~\citeyear{Merrit09}).

It is well established that massive black holes regularly
populate the centers of massive galaxies (e.g. Kormendy \&
Richstone~\citeyear{Kormen95}, Magorrian et al.~\citeyear{Magorr98}, Ferrarese
\& Merritt~\citeyear{Ferrar00}, Gebhardt et al.~\citeyear{Gebhar00}, Marconi
\& Hunt~\citeyear{Marcon03}, H\"aring \& Rix~\citeyear{Haring04}), with the BH
mass increasing proportionally for more massive host galaxies. At the same
time, the frequency of nuclear star clusters in the center of
massive galaxies decreases with increasing host galaxy mass (e.g. C{\^o}t{\'e}
et al.~\citeyear{Cote06}). This is typically interpreted as the effect of
dynamical heating from supermassive black holes in centers of massive
galaxy spheroids, which can contribute to evaporate a compact stellar system
(e.g. Bekki \& Graham~\citeyear{Bekki10}, Antonini ~\citeyear{Antoni13}).

Recent literature suggests that BHs and NCs often coexist in galaxy
spheroids of intermediate mass (e.g. Gonzales Delgado et
al.~\citeyear{Gonzal08}, Graham \& Spitler~\citeyear{Graham09}, Seth et
al.~\citeyear{Seth08} \& \citeyear{Seth10}, Graham~\citeyear{Graham12}). For host
galaxy masses of $10^{10}\msun$ and below, BH masses are comparable to or
smaller than the mass of the NC. Above that host galaxy mass, BHs
generally dominate the mass budget (Graham~\citeyear{Graham12}, Graham \&
Scott~\citeyear{Graham13}) and exceed the masses of NCs.

The host galaxy masses suggested for UCD progenitors are in the
range $10^8$--$10^{10}\msun$ (e.g. Bekki
et al.~\citeyear{Bekki03}, Pfeffer \& Baumgardt~\citeyear{Pfeffe13}). In this galaxy mass range, central BHs
are typically less massive than NCs. If UCDs were identical to such
NCs that then got tidally stripped, their BHs (if existing) would be
expected to have modest, and not dramatic, effects on their internal
dynamics. This is qualitatively consistent with the 50\% elevated M/L
ratios found on average for UCDs. In this context it is interesting to
note that the formation of massive BHs in compact stellar systems
(CSSs) may not necessarily require that CSS to be embedded in a deep
potential well: Leigh et al.~(\citeyear{Leigh13}) discuss the growth of
BHs via accretion in massive primordial clusters and find that
Bondi-Hoyle accretion may create $10^5$--$10^6\msun$ solar mass BH in
systems of total mass $\sim 10^7\msun$, similar to UCD masses. They
explicitly predict that the formation of such BHs in massive clusters
could influence the measured dynamical M/L ratio. Going one step
further, if UCDs were to represent `hyper-compact stellar systems'
bound to recoiling super-massive black holes that were ejected from
the centres of massive galaxies (Merritt et al.~\citeyear{Merrit09}), they
would contain a black hole with a mass comparable to, or even
exceeding that of, the stellar component. 

In a first attempt to search for BH evidence in a UCD, Frank et
al.~(\citeyear{Frank11}) presented a spatially resolved velocity
dispersion profile of the Fornax cluster UCD3, using seeing restricted
FLAMES IFU observations obtained with the VLT. In their study they did
not find evidence for a rise of the projected radial velocity
dispersion $\sigma$ towards its center at a level beyond that expected
from a simple mass-follows-light model. They derive an upper BH mass
limit of about 5\% of the total UCD mass for UCD3. Assuming their best
fitting mass-follows-light model, they found a dynamical M/L for UCD3
in excellent agreement with the one derived from previous global
(i.e. single aperture) spectroscopy, which in turn is fully
consistent with a normal stellar population (Chilingarian et
al.~\citeyear{Chilin11} and this paper).

For less massive compact stellar systems like the Local Group objects
omega Centauri, G1 (Mayall II, in M31), and lower mass globular
clusters, observational evidence both in favour of and against central
intermediate-mass black holes has been presented and discussed
(e.g. Gehbardt et al.~\citeyear{Gebhar02}, Ulvestad et
al.~\citeyear{Ulves07}, Maccarone et al.~\citeyear{Maccar07} \&
\citeyear{Maccar08}, Noyola et al.~\citeyear{Noyola10}, Anderson \&
van der Marel~\citeyear{Anders10}, Cseh et al.~\citeyear{Cseh10},
L\"{u}tzgendorf et al.~\citeyear{Luetzg13}, Feldmeier et
al.~\citeyear{Feldme13}, Lanzoni et al.~\citeyear{Lanzon13}). For any
of the above studies, the upper mass limits for central BHs do not
exceed a few percent of the total cluster mass.

\subsection{This paper}

In this paper, we estimate hypothetical central BH masses for those
UCDs that have dynamical M/L ratios above the expectation from
canonical stellar population models. For this we assume that a central
BH is responsible for the seemingly elevated M/L ratio. To allow a
self-consistent BH mass estimate, we perform in Sect.~\ref{recalc} a
homogeneous recalculation of UCD dynamical masses using available
literature information, including spectroscopic aperture correction
and updated distance moduli. In Sect.~\ref{BH} we estimate central BH
masses for those UCDs for which the dynamical M/L ratio is above the
stellar population M/L ratio. In Sect.~\ref{discussion} we discuss our
findings in the context of UCD formation scenarios and the well known
scaling relations between BHs and their hosts. We finish this paper in
Sect.~\ref{summary} with Conclusions and Outlook, including a
calculation of the projected velocity dispersion profile expected for
the two UCDs with the most massive predicted BHs.

\section{Homogeneous recalculation of UCD dynamical masses}
\label{recalc}
In order to allow self-consistent BH mass estimates, we first perform
a homogeneous recalculation of literature dynamical M/L measurements
of compact stellar systems with dynamical masses $\gtrsim 2 \times
10^6\msun$ outside the Local Group. We refer to those sources as UCDs
in the following. We also use this recalculation to estimate the
amount of systematic uncertainties involved in the dynamical
modelling, by comparing our dynamical mass estimates with previous
literature results. The environments included are: Centaurus A (NGC
5128, CenA in the following; Rejkuba et al.~\citeyear{Rejkub07} \&
Taylor et al.~\citeyear{Taylor10}), the Virgo cluster (Ha\c{s}egan et
al.~\citeyear{Hasega05}, Evstigneeva et al.~\citeyear{Evstig07},
Chilingarian \& Mamon~\citeyear{Chilin08}), and the Fornax cluster
(Hilker et al.~\citeyear{Hilker07}, Mieske et al.~\citeyear{Mieske08},
Chilingarian et al.~\citeyear{Chilin11}). The full sample comprises 53
UCDs, for 49 of which we remodel the internal dynamics. These are
listed in Table~\ref{tableall}. The four remaining UCDs, all in
CenA, are excluded due to inconsistent information on structural
parameters in the literature (see Sect.~\ref{exclude} below).

When investigating the internal dynamics of marginally resolved
extragalactic sources, it is crucial to correctly model the effects of
seeing and aperture on the resulting observed velocity dispersion
$\sigma_{\rm obs}$. Typical intrinsic sizes of those systems are a few to
a few tens of parsec (Mieske et al.~\citeyear{Mieske08}, Misgeld \&
Hilker~\citeyear{Misgel11}), corresponding to angular diameters of a
few tenths of arcseconds. This approaches typical aperture sizes in
ground-based spectroscopy, such that aperture corrections are
necessary to correctly derive the dynamical mass of UCDs (Hilker et
al.~\citeyear{Hilker07}, Evstigneeva et al.~\citeyear{Evstig07},
Taylor et al.~\citeyear{Taylor10}). Generally it holds for the
central velocity dispersion $\sigma_0$, the measured value
$\sigma_{\rm obs}$ and the global dispersion $\sigma_{\rm global}$ that
$\sigma_0 > \sigma_{\rm obs} > \sigma_{\rm global}$. Typically, the observed
$\sigma_{\rm obs}$ is closer to $\sigma_{\rm global}$ than to $\sigma_0$ since
the intrinsic object sizes are mostly below the size of spectroscopic
aperture and seeing FWHM.

Input for our remodelling is the measured velocity dispersion $\sigma_{\rm obs}$, the
surface brightness profile (mostly from HST), the seeing FWHM of the
observations, the extraction area used in the spectroscopic setup, and
an updated distance modulus. We use a distance modulus (m-M)=27.88 mag
for CenA (Taylor et al.~\citeyear{Harris10}), 31.09 mag for Virgo (Mei
et al.~\citeyear{Mei07}), and 31.5 mag for Fornax (Blakeslee et
al.~\citeyear{Blakes09}). See the list of all sources in
Table~\ref{tableall}.

\subsection{Adopted surface brightness profile parameters}
\label{SBprofile}

Surface brightness profiles as derived in the literature are defined
by the density profile used (e.g. King~\citeyear{King62},
King~\citeyear{King66}, S{\'e}rsic). For the most often used King (1962)
and King (1966) profiles, structural parameters consist of the tidal
radius $r_t$, core radius $r_c$ and the King concentration parameter
$c$ defined by

\begin{equation}
c=log(r_t/r_c)
\label{kingc}
\end{equation}

The so-called generalised King profile is given by

\begin{equation}
I(R) = I_0 \left[\frac{1}{(1+(r/r_c)^2)^{\frac{1}{\alpha}}} -
\frac{1}{(1+(r_t/r_c)^2)^{\frac{1}{\alpha}}} \right]^{\alpha},
\label{king}
\end{equation}

where $\alpha$ is an additional parameter describing the shape of the
surface density profile. With $\alpha=2$ this becomes the standard
King~(\citeyear{King62}) surface brightness profile.

In the fitting of King~(\citeyear{King62}) profiles, only two of the
three quantities $r_h$, $r_c$, and $c$ are direct fit outputs. The
third quantity is derived from the other two. Most authors fit for the
two King parameters core radius $r_c$ and concentration $c$, and then
derive the half-light radius $r_h$ analytically. To verify that the
literature values are self-consistent, we rederived $r_h$ from the
other model parameters and compared them to the published literature
values. For some sub-samples (see next subsection), slight corrections
needed to be applied to obtain self-consistent values of $r_h$. After
these corrections were applied, the average ratio between both $r_h$
values was 1.00 with an object-to-object RMS of 0.035.

In Table~\ref{tableall} we indicate for these 49 sources the surface
brightness profile parameters adopted in this study. We use as
distance moduli for the three environments the values cited in the
previous subsection.

\subsubsection{Notes on individual objects}
\label{special}
In the following we give some notes on the adopted surface brightness
profiles of various sub-samples

\vspace{0.1cm}
\noindent {\bf Corrections to $r_c$ from Ha\c{s}egan et
  al.~(\citeyear{Hasega05}):} For the Virgo UCDs analysed by
Ha\c{s}egan et al.~(\citeyear{Hasega05}), the authors fit for the
half-light radius $r_h$ and concentration $c$, and then determine the
King radius via an analytical formula adopted from McLaughlin et
al.~(\citeyear{McLaug00}). We note however that the corresponding
formula as given by McLaughlin et al.~(\citeyear{McLaug00}) Appendix
B4, refers to the King scale radius $r_0=(9 \sigma /(4\pi G
\rho_0))^{1/2}$, which is not identical to the core radius $r_c$. For
small concentrations $r_0$ differs systematically from $r_c$ (defined
as radius where the surface brightness drops to 50\% of the central
value), as also noted in footnote 3 of McLaughlin et
al.~(\citeyear{McLaug00}). The core radius $r_c$ and King radius $r_0$
were recalculated by us based on the half light radii $r_h$ and the
concentration $c$ as given in Table 7 of Ha\c{s}egan et
al.~(\citeyear{Hasega05}). The resulting $r_0$ values show excellent
agreement with the values given under the misleading label $r_c$ by
Ha\c{s}egan et al.~(\citeyear{Hasega05}). The recalculated values of
$r_c$ are between 0\% and 10\% smaller than the values quoted by
Ha\c{s}egan et al.~(\citeyear{Hasega05}). We list our corrected
$r_c$ values in Table ~\ref{Hasegan_sizes}.

\begin{table}
\caption{List of the six Virgo UCDs from Ha\c{s}egan et al.~(\citeyear{Hasega05}) for which we adopted corrected King core radii $r_c$. The values listed as core radii $r_c$ in Ha\c{s}egan et al.~(\citeyear{Hasega05}) correspond to King scale radii $r_0$, which differ from $r_c$ especially for low concentrations $c$ (McLaughlin et al.~\citeyear{McLaug00}, footnote 3). The corrections are between 0\% and 10\%. Note that the values r$_{\rm c,Hasegan}$ and $r_h$ are given for our adopted distance modulus 31.09 mag to Virgo (Mei et al.~\citeyear{Mei07}), which is slightly larger than the distance modulus of 31.03 mag originally adopted in Ha\c{s}egan et al.~(\citeyear{Hasega05}).}
\label{Hasegan_sizes}
\begin{center}
\begin{tabular}{rrrrrrrr}
\hline \hline
\noalign{\smallskip}
Name & r$_{\rm c,Hasegan}$ [pc] & $r_{\rm c,corrected}$ [pc]& c & r$_{\rm h}$ [pc]\\
  \noalign{\smallskip}
\hline
\noalign{\smallskip}
 H8005    &   14.0  &     13.1  &    1.3 &  29.5   \\
  S314    &   0.84   &     0.81   & 1.7 &    3.32    \\
  S417   &     8.1  &      7.4  & 1.19 &     14.7   \\ 
  S490   &   0.70  &      0.72  & 1.84 &    3.74   \\
  S928    &   13.9  &     12.6  &  1.13  & 23.85   \\
  S999    &   13.2    &   11.7  &  1.05 &   20.69 \\  
\end{tabular}
\end{center}
\end{table}

\vspace{0.2cm}

\noindent {\bf Sources not parametrised by a single King profile:} For
six objects in Table~\ref{tableall} marked with an asterisk $^*$ the
surface brightness profile according to the literature sources is not
adequately described by a single King (\citeyear{King62} or
\citeyear{King66}) profile. The following profiles from the literature
were adopted for those sources:
\begin{itemize}
\item M59cO (Chilingarian \& Mamon~\citeyear{Chilin08}): double S{\'e}rsic
  profile with n=1: r$_{\rm h,in} =13 \pm 1$ pc, m$_{\rm tot,in} = 19.78$ mag,
  r$_{\rm h,out} = 50 \pm 2$ pc, m$_{\rm tot,out} = 18.82$ mag; B-band
  luminosity M$_B$=$-$12.34 mag.

\item VUCD7 (Evstigneeva et al.~\citeyear{Evstig07}): double S{\'e}rsic profile. S{\'e}rsic core with n=2.2, r$_{\rm h}=$9.9 pc $\mu_{\rm eff}$=16.15 mag; 
S{\'e}rsic envelope with n=1.4,  r$_{\rm h}=$231 pc, $\mu_{\rm eff}$=22.31 mag.

\item F19 (Mieske et al.~\citeyear{Mieske08}): composite profile King
  + S{\'e}rsic. King ~(\citeyear{King66}) core $r_c$=3.77 pc, c=1.52, $\mu_0$=15.74 mag,
  alpha=2, S{\'e}rsic envelope with n=1.3, $r_h$=111.6 pc, $\mu_{\rm \rm
    eff}=21.13$ mag.

\item UCD5 (Evstigneeva et al.~\citeyear{Evstig07}): composite profile
  King + S{\'e}rsic: King ~(\citeyear{King62}) core $\mu_0$=15.13 mag, $r_t$=33.6 pc,
  $r_c$=4.36 pc. S{\'e}rsic envelope $\mu_{\rm \rm eff}$=24.04 mag, n=6.88,
  r$_h$=147.4 pc.

\item F8 (Chilingarian et al. \citeyear{Chilin11}): Single S{\'e}rsic
  profile with n=1.2, $r_h$=6.7 pc.

\item F2 (Chilingarian et al. \citeyear{Chilin11}): Single S{\'e}rsic
  profile with n=4.9, $r_h$=14.7 pc.

\end{itemize}

\vspace{0.1cm}
\noindent {\bf Sources with King exponent $\alpha \neq 2$:}
For the following eight sources, the literature quotes generalised King surface brightness profiles (Eq.~\ref{king}) with exponent $\alpha$ different from 2. Those are: F1
[$\alpha=1.23$], F22 [$\alpha=1.10$], F24 [$\alpha=3.32$], F7
[$\alpha=2.77$], UCD1 [$\alpha=0.74$], UCD12a [$\alpha=1.30$], VUCD1
[$\alpha=3.74$], VUCD3 [$\alpha=0.62$].

\vspace{0.2cm}
\noindent {\bf Updated half-light radii for Fornax UCDs Fxx:} For the
Fornax Fxx sources (first quoted in Mieske et al.~\citeyear{Mieske08})
we use updated half-light radii determined directly from the King
(\citeyear{King62}) model fits. The previously published values for
these sources (Table 1 of Mieske et al.~\citeyear{Mieske08}) were
determined from a curve of growth analysis that is independent of the
King profile fit.

\subsubsection{Objects excluded from the sample}
\label{exclude}
As mentioned above, there are four CenA UCDs for which we find
substantial disagreement between literature (Holland et
al.~\citeyear{Hollan99}) and recalculated $r_h$. They are HCHH99-15,
HCHH99-18 (dynamical mass determined in Rejkuba et
al.~\citeyear{Rejkub07} and remodelled in Mieske et
al.~\citeyear{Mieske08}), and GC 217 and 242 from Taylor et
al.~(\citeyear{Taylor10}), which are HCHH99-16 and HCHH99-21 in the
notation of Holland et al.~(\citeyear{Hollan99}). We list in
Table~\ref{HCHH99_sizes} the literature set of $r_c$, $r_h$ and
concentration, and indicate the 2D half-light radius $r_h$ that is
actually obtained from $r_c$ and $c$ as given in Holland et
al.~(\citeyear{Hollan99}). The two values of $r_h$ differ by a factor
of 2-4.  Due to this substantial disagreement we exclude these
four objects from our analysis.

\begin{table*}
\caption{List of the four UCDs from our original 53 UCD sample for
  which we find substantial disagreement (factor of 2-4) between literature 2D
  half-light radius r$_{\rm h,HCHH99}$, and r$_{\rm h,this paper}$
  recalculated using the King profile parameters $r_c$ and $c$ as given in
  the same literature source. These are four CenA sources with the
  structural parameters derived in Holland et al. (\citeyear{Hollan99};
  HCHH99). Sizes r$_h$, $r_c$ are in pc. \label{HCHH99_sizes}}
\begin{center}
\begin{tabular}{rrrrrr}
\hline \hline
\noalign{\smallskip}
Name & r$_{\rm h,HCHH99}$ & $r_c$& c & r$_{\rm h,this paper}$ &Literature\\
 \noalign{\smallskip}
\hline
\noalign{\smallskip}
HCH99-C16 (GC 217)&      11.9  &  0.80 & 1.60 &  2.63& Taylor+10\\
HCH99-C21 (GC 242)&       7.0  &  2.52 & 0.80 &  3.40& Taylor+10\\
HCH99-C15 &   5.9  &  1.60 & 1.00  & 2.72&Rejkuba+07, Mieske+08\\
HCH99-C18 &  13.7  &  1.20 & 1.50  & 3.55&Rejkuba+07, Mieske+08\\
\end{tabular}
\end{center}
\end{table*}

\subsection{Procedure for determining the dynamical UCD mass}
\label{procedure}

We recalculate the dynamical mass of all UCDs under the assumption
that mass follows light. For this we follow a similar procedure as
outlined in Hilker et al.~(\citeyear{Hilker07}) and Mieske et
al.~(\citeyear{Mieske08}). An input for this calculation is the
stellar population mass $M_{\rm pop}$ of each UCD, which is obtained by
multiplying the stellar population mass-to-light ratio $M/L_{\rm pop}$ with
the luminosity.

\begin{equation}
M_{\rm pop} = M/L_{\rm pop} \times 10^{0.4\times(-M_V+4.83)}
\label{mphot}
\end{equation}

For $M_V$ we use the V-band apparent magnitudes from each literature
source and their updated distance moduli (see above). For $M/L_{\rm pop}$
we adopt the mean of stellar population model predictions from
Maraston et al. (\citeyear{Marast05}) and Bruzual \& Charlot
(\citeyear{Bruzua03}), evaluated at the object's metallicity [Fe/H],
for an age of 13 Gyr and solar alpha abundance. We find that
$M/L_{\rm pop}$ as a function of [Fe/H] is well described by an
exponential relation of the form

\begin{equation}
M/L_{\rm pop}= 0.5\times(1.98\times e^{(1.245[\mathrm{Fe/H}])}+1.72+2.7\times e^{(1.09[\mathrm{Fe/H}])}+1.8)
\label{mlstellar}
\end{equation}

which we adopt in the following (see also Fig.~\ref{massML}, and
Mieske et al.~\citeyear{Mieske08}, Dabringhausen et
al.~\citeyear{Dabrin08}).

Using the stellar population mass of the UCDs, we set up a model to predict
the observed sigma $\sigma_{\rm obs,pop}$, assuming that the mass
distribution follows the light. This value is then compared to the
spectroscopically measured velocity dispersion $\sigma_{\rm obs}$. The
square of the ratio of the two $\sigma$ values gives the deviation
between dynamical and stellar population mass, yielding the dynamical
mass. The individual steps are:

\begin{enumerate}    

\item We assume that the UCDs are spherically symmetric and have
  isotropic velocity distributions.  The 2-dimensional surface density
  profile as quantified by the surface brightness parameters from
  Table 2 - typically King core radius and concentration -- is
  deprojected by means of Abel's integral equation (equation 1B-59 of
  Binney \& Tremaine~\citeyear{Binney87}) into a 3-dimensional density
  profile $\rho (r)$.

\item The cumulated mass function M(<r), the potential energy $\phi(r)$
  and the energy distribution function f(E) (equation 4-140b of Binney
  \& Tremaine~\citeyear{Binney87}) are calculated from the
  3-dimensional density profile. 

\item We sample the 6D phase space density distribution of the UCDs by creating an
  N-body representation of each UCD. In total 10$^5$ particles are
  distributed in radius according to the density profile $\rho (r)$ 
  calculated above and each particle is assigned a velocity according
  to the distribution function f(E). We then calculate x, y, z
  positions and vx, vy and vz velocities from the radii and velocities
  assuming spherical symmetry and an isotropic velocity dispersion.

\item The influence of seeing is modelled by assuming that the light
  from each particle is distrbuted as a 2D Gaussian whose full width
  at half maximum (FWHM) corresponds to the observed seeing. For each
  particle the fraction of the light that falls within the
  spectroscopic aperture is calculated.

\item The fraction of light is used as weighting factor for the
  velocities. The weighted velocity contributions are then used to
  calculate the expected velocity dispersion $\sigma_{\rm \rm obs,pop}$
  for the stellar population mass $M_{\rm pop}$ of the UCD (as
  determined from its absolute magnitude and metallicity, assuming a
  13 Gyr stellar population, see Table 2).

\item The dynamical mass is calculated from $\M_{\rm dyn}=\M_{\rm pop} \times
  \frac{\sigma_{\rm \rm obs}^2}{\sigma_{\rm \rm obs,pop}^2}$, with
  $\sigma_{\rm \rm obs}$ being the spectroscopically observed velocity
  dispersion. At this step, we assume that the gravitational mass is
  distributed like the luminous mass. We note that the result for
  $\M_{\rm dyn}$ does not depend on the particular choice of the reference
  mass for which $\sigma_{\rm \rm obs,pop}$ is calculated. 

\end{enumerate}

\begin{figure*}[]
\begin{center}
\includegraphics[width=8.6cm]{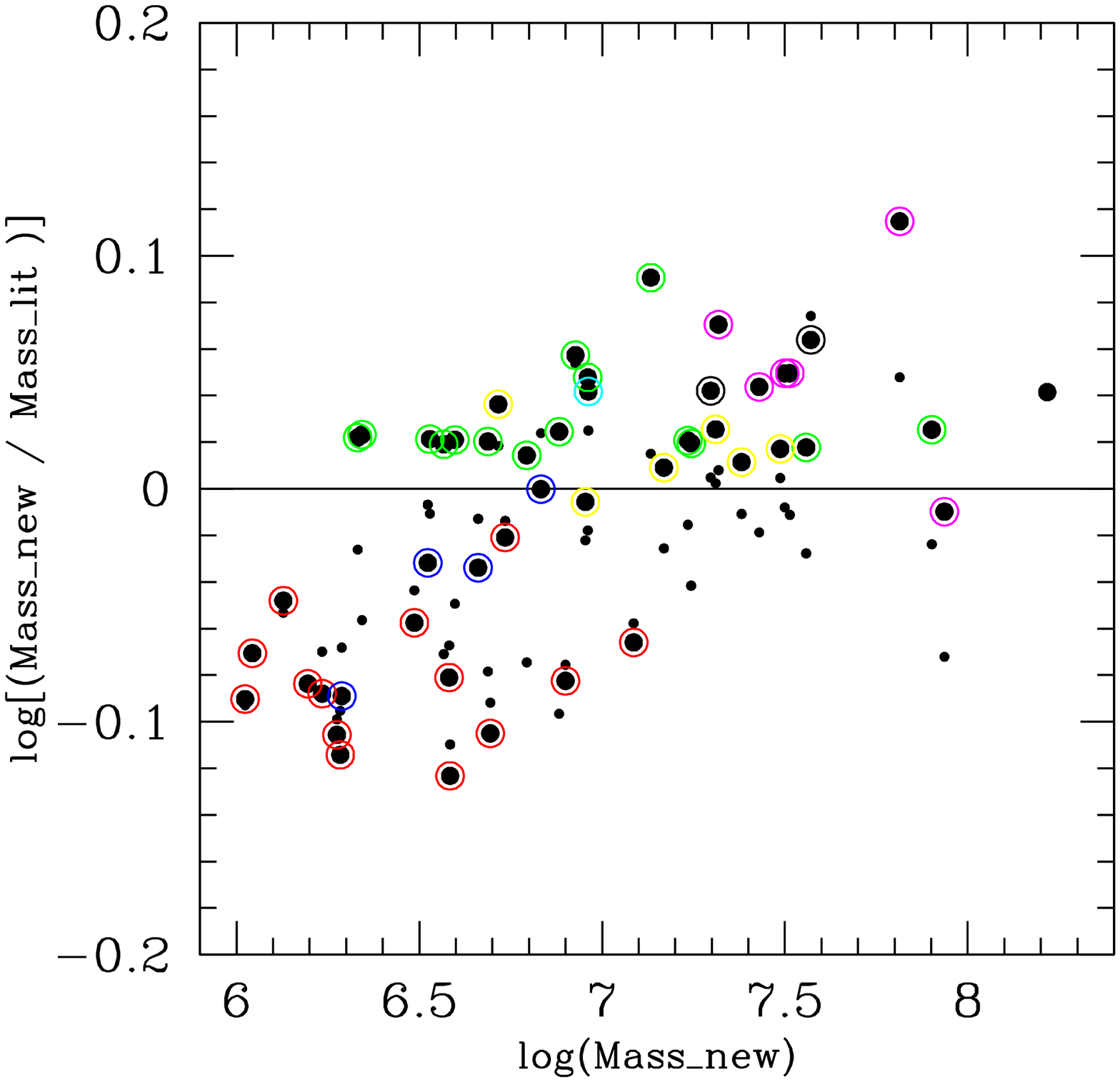}
\includegraphics[width=8.6cm]{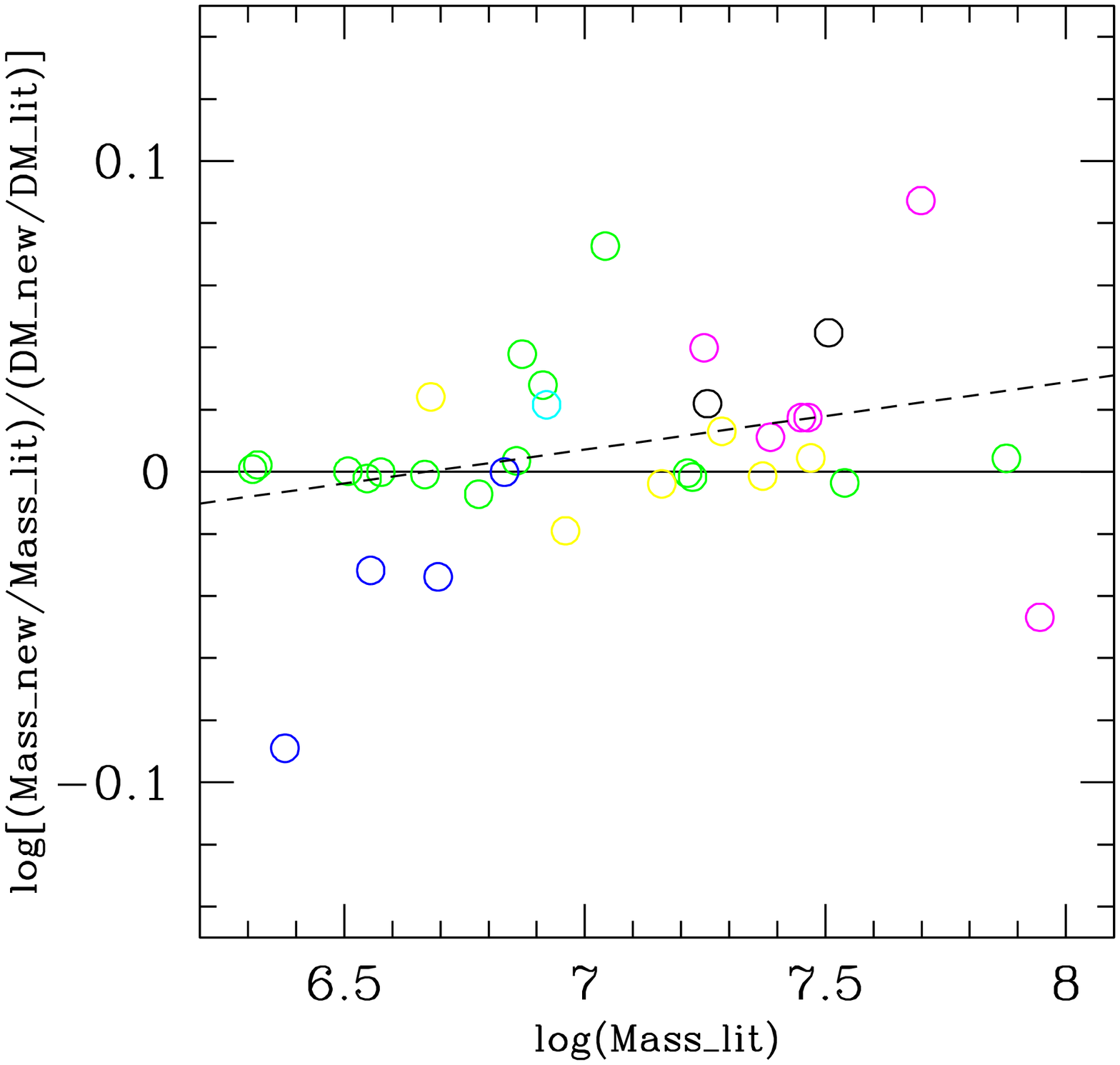}

  \caption{{\bf Left panel}: The ratio between remodelled and previous
    literature estimates of dynamical masses of UCDs is plotted vs
    their remodelled mass. The original literature source is indicated
    by the colour coding of the circles around the data points. See
    Sect.~\ref{comparesec} for details on the colour coding. For all the
    data points, we have recalculated the dynamical mass in this
    paper. Small dots (without a circle) indicate the ratio between
    the remodelled and literature dynamical mass-to-light ratio
    $M/L_{\rm dyn}$. {\bf Right panel:} The ratio between remodelled
    and literature mass, cleaned of the effect of update in distance
    modulus, and without the CenA sample of Taylor et
    al.~(\citeyear{Taylor10}), is plotted vs. the literature mass
    estimate.}
\label{compare}
\end{center}
\end{figure*}

\subsection{Comparison of new and literature model masses}
\label{comparesec}
Based on the dynamical UCD masses derived above under the assumption
that mass-follows-light we use the V-band apparent magnitudes from
each literature source and their updated distance moduli to calculate
the dynamical M/L ratios.

In Fig.~\ref{compare} we plot the newly obtained dynamical masses
vs. the ratio between this new estimate and the literature
estimate. Note that all literature values were already based on a
similar modelling as in this paper. We indicate as an aid for the eye
as solid line the identity. The data points are colour coded according
to the literature source of their original measurement. Blue circles
refer to CenA UCDs originally from Rejkuba et
al. (\citeyear{Rejkub07}) with a first remodelling presented in Mieske
et al. (\citeyear{Mieske08}). Yellow circles refer to Virgo UCDs
originally from Ha\c{s}egan et al. (\citeyear{Hasega05}) with a first
remodelling presented in Mieske et al. (\citeyear{Mieske08}). Green
circles refer to Fornax UCDs from Chilingarian et
al. (\citeyear{Chilin11}), which itself is based on a remeasurement of
spectroscopic data from Mieske et al. (\citeyear{Mieske08}) and a
remodelling performed in Chilingarian
et. al. (\citeyear{Chilin11}). The cyan circle refers to one Fornax
UCD with designation F12 from Mieske et al. (\citeyear{Mieske08}). Red
circles refer to UCDs in CenA from Taylor et
al. (\citeyear{Taylor10}).  Magenta circles refer to Virgo UCDs from
Evstigneeva et al. (\citeyear{Evstig07}). Black circles indicate
Fornax UCDs from Hilker et al. (\citeyear{Hilker07}). The object
without a circle is Virgo UCD M59cO, from Chilingarian \& Mamon
(\citeyear{Chilin08}).

The median ratio between our newly modelled dynamical mass estimates
and the literature values is 1.04 with an RMS scatter of
0.13. Similarly, the ratio between the new and literature $M/L_{\rm
  dyn}$ is 0.95 with an RMS scatter of 0.13. The opposite sense of
difference in mass and mass-to-light ratio is mainly due to the
updates in distance modulus for the Fornax and Virgo sources. They are
between 3\% and 8\% larger than the distances adopted in the
respective articles (see Table 1). Total masses increase linearly with
distance, while M/L ratios decrease linearly since L scales
quadratically with distance. The RMS scatter of 0.13 is not negligible
when investigating individual sources in detail. Its value gives an
indication of the systematic uncertainties in the determination of
dynamical masses from integrated velocity dispersions, and should be
considered for the total error budget of such measurements.

In the following we briefly discuss the one sample with a notable
systematic offset between literature and remodelled mass, which is the
CenA sample from Taylor et al. (\citeyear{Taylor10}).

\subsubsection{CenA UCDs from Taylor et al. (2010).}
  
  The dynamical masses (and mass-to-light ratios) derived by Taylor et
  al. are on average 17\% above our remodelled values, with an
  RMS of 6\%. The fact that the average offset is larger than the
  scatter indicates a systematic relative bias in the mass
  calculation. Taylor et al. perform a modelling analogous to Hilker
  et al. (\citeyear{Hilker07}) and Mieske et al. (\citeyear{Mieske08}), to
  find the UCD model which represents best the measured velocity
  dispersion, given the spectroscopic aperture, seeing, and intrinsic
  light distribution of the source. However, unlike other studies,
  their finally quoted dynamical mass is {\it not} the total mass of
  the best fitting model. Instead, it is the dynamical mass derived
  from the modelled central velocity dispersion $\sigma_0$, via
  equation 3 of their paper: $M_{\rm dyn} = \frac{\beta \times \sigma_0^2
    \times r_h}{G}$. They adopt a value of $\beta=7.5$ for all their
  sources. They follow this approach of using $\sigma_0$ in order to
  contrast their measurement with literature data on other hot stellar
  systems which typically analyse scaling relations with $\sigma_0$.

  Generally, $\beta$ is a scaling factor that varies strongly
  depending on the surface brightness profile. The typical ranges are
  between 5-10, see the literature review in Taylor et
  al. (\citeyear{Taylor10}). Thus, one expects an object-to-object
  scatter when comparing the results of Taylor et al. - which assume a
  fixed $\beta$ - to our remodelled masses which use the individual
  surface brightness profiles. The additional, systematic 17\% average
  offset would be removed by adopting $\beta \sim 6.5$ instead of 7.5,
  which is still in the range of typical literature values for
  $\beta$. For example, Cappellari et al. (\citeyear{Capell06}) find
  $\beta=5.0 \pm 0.1$ for the central regions of early type galaxies
  and that $\beta$ varies by a factor of 2.5 amongst the objects in
  their sample. We thus conclude that the choice of a fixed $\beta$ is
  the most likely reason for the offset and scatter in the dynamical
  mass comparison between Taylor et al. (\citeyear{Taylor10}) and our
  remodelled data, noting that also in our analysis we make 
  a simplifying assumption, namely an isotropic velocity distribution.

\subsubsection{An overall trend?}

In Fig.~\ref{compare} left panel one notices an apparent trend in
the sense that the ratio between new and literature mass increases
with mass. To investigate the significance of this, we clean the
sample in the following way, and replot it in the right panel. A) We
remove the CenA sample from Taylor et al (2010), since this sample
exhibits a systematic overall offset discussed above. B) We
remove from the newly determined masses the effect of distance modulus
update, reducing the mass estimates for the Fornax and Virgo sources
between 3 and 8 \%. C) We plot on the x-axis the literature mass
estimate instead of the new mass estimate. 

We then fit a linear relation to the data points. The resulting fit as
indicated in the right panel of Fig.~\ref{compare} yields a weak
correlation with a slope of 0.021 $\pm$ 0.012 in log-log space between
the mass ratio and the mass, which is significant at the 1.8$\sigma$
level. This corresponds to a 5\% effect per dex in mass, or a
range of $\pm 4\%$ between mean and extreme masses of the sample. In
the context of our study this is negligible.

\vspace{0.2cm}

\begin{longtab}
\small
\begin{landscape}
\begin{longtable}{rrrrrrrrrrrrrrr}
\caption{Recalculated dynamical masses and further properties of UCDs in Centaurus A, Virgo, and Fornax. $\Psi$ is defined as $\Psi = (M/L)_{\rm dyn}/(M/L)_{\rm pop}$.  The second-to-last column denotes the literature source of the original measurement of M$_{\rm dyn,remod}$, and the origin of the [Fe/H] value. {\bf lit=1:} Cen A UCDs from Mieske et al. (\citeyear{Mieske08}), based on original measurements presented in Rejkuba et al. (\citeyear{Rejkub07}). {\bf lit=2:} Virgo UCDs from Mieske et al. (\citeyear{Mieske08}), based on original measurements presented in Ha\c{s}egan et al (\citeyear{Hasega05}). {\bf lit=3:} Virgo UCDs from Evstigneeva et al. (\citeyear{Evstig07}). {\bf lit=3.2:} The Virgo UCD M59cO from Chilingarian \& Mamon (\citeyear{Chilin08}). {\bf lit=3.5:} Fornax UCDs from Hilker et al. (\citeyear{Hilker07}). {\bf lit=7:} CenA UCDs from Taylor et al. (\citeyear{Taylor10}), largely based on a re-reduction of data published in Rejkuba et al. (\citeyear{Rejkub07}). {\bf lit=8:} Fornax UCDs from Chilingarian et al. (\citeyear{Chilin11}), doing a remeasurement and re-modelling of spectroscopic data from Mieske et al. (\citeyear{Mieske08}). For objects marked with an asterisk $^*$, no unique King profile fit is available in the literature. Composite or single Sersic fits are adopted, see Sect.~\ref{SBprofile} for details. For VUCD7 and F19, the indicated King profile parameters r$_c$ and $c$ are for the core component. For F19, the small error bar of 0.3 km/s for the measured velocity dispersion is due to the very high S/N spectrum obtained for this object, combined with the usage of the full spectral fitting technique (Chilingarian et al.~\citeyear{Chilin11}).\label{tableall}}\\
\hline \hline
\noalign{\smallskip}
Name & M$_{\rm dyn}$ [10$^6 \msun$] & [Fe/H] & M$_V$ & M/L$_{\rm dyn}$ & M$_{\rm pop}$ [10$^6 \msun$]& $\Psi$ & $\frac{M_{\rm dyn,this paper}}{M_{\rm dyn,literature}}$ & $\sigma$ [km/s] & r$_{\rm h}$ [pc] & r$_c$ [pc] & $c$ & lit & dist [Mpc] & BHfrac\\
 \noalign{\smallskip}
\hline
\noalign{\smallskip}

\endfirsthead
\caption{continued.}\\
\hline \hline
\noalign{\smallskip}

Name & M$_{\rm dyn}$ [10$^6 \msun$]& [Fe/H] & M$_V$ & M/L$_{\rm dyn}$ & M$_{\rm pop}$ [10$^6 \msun$]& $\Psi$ & $\frac{M_{\rm dyn,this paper}}{M_{\rm dyn,literature}}$ & $\sigma$ [km/s] & r$_{\rm h}$ [pc] & r$_c$ [pc] & $c$ & lit & dist [Mpc] & BHfrac\\

\noalign{\smallskip}
\hline
\noalign{\smallskip}
\endhead
\noalign{\smallskip}
\hline
\endfoot
M59cO$^*$ & 160 $\pm$   34 &  -0.03 & -13.26 &  9.6 $\pm$  2.4 & 69 &  2.38 $\pm$ 0.59 &   1.10 &    48 $\pm$     5 &     32 &  --- &  --- &  3.2 &  14.9 & 0.31$^{+0.15}_{-0.12 }$ \\
VUCD7$^*$ &   86 $\pm$   19 &  -0.66 & -13.61 &  3.6 $\pm$ 0.93 & 68 &  1.27 $\pm$ 0.33 & 0.978 &    38.0 $\pm$   4.1 &    105 &  --- &  --- &    3 &  16.5 & 0.027$^{+0.03}_{-0.027 }$ \\
F19$^*$ &   80 $\pm$  1.9 &  -0.19 & -13.49 &  3.7 $\pm$ 0.49 & 77 &  1.03 $\pm$ 0.13 &  1.06 &    24.8 $\pm$   0.3 &   90.6 &  3.7 &  1.7 &    8 &    20 & 0.003$^{+0.014}_{-0.003 }$ \\
VUCD3 &   65 $\pm$  4.6 &  -0.011 & -12.75 &  6.1 $\pm$  1.5 & 44 &  1.49 $\pm$ 0.37 &   1.30 &    42.1 $\pm$   1.5 &   21.6 &  2.3 &  2.1 &    3 &  16.5 & 0.078$^{+0.062}_{-0.057 }$ \\
UCD1 &   37 $\pm$  2.3 &  -0.67 & -12.19 &  5.8 $\pm$ 0.97 & 18 &  2.04 $\pm$ 0.34 &  1.16 &    32.0 $\pm$     1.0 &   23.5 &  1.9 &  2.3 &  3.5 &    20 & 0.15$^{+0.058}_{-0.04 }$ \\
F24 &   36 $\pm$  2.5 &  -0.67 & -12.37 &  4.8 $\pm$ 0.72 & 22 &  1.68 $\pm$ 0.25 &  1.04 &    29.1 $\pm$     1.0 &   30.9 &  3.1 &  3.2 &    8 &    20 & 0.11$^{+0.044}_{-0.038 }$ \\
VUCD5 &   33 $\pm$  3.7 &  -0.36 & -12.49 &  3.9 $\pm$ 0.77 & 28 &  1.17 $\pm$ 0.23 &  1.12 &    28.2 $\pm$   1.6 &   19.2 &  7.1 &  1.4 &    3 &  16.5 & 0.042$^{+0.055}_{-0.042 }$ \\
VUCD1 &   32 $\pm$  3.1 &  -0.76 & -12.43 &  3.9 $\pm$ 0.81 & 22 &  1.44 $\pm$  0.3 &  1.12 &    34.1 $\pm$   1.7 &   12.1 &  4.6 &  1.9 &    3 &  16.5 & 0.11$^{+0.074}_{-0.065 }$ \\
S417 &   31 $\pm$    3 &  -0.70 & -11.84 &  6.6 $\pm$  1.5 & 13 &  2.35 $\pm$ 0.53 &  1.04 &    30.5 $\pm$   1.5 &   14.7 &  8.1 &  1.2 &    2 &  16.5 &  0.4$^{+0.17}_{-0.15 }$ \\
VUCD4 &   27 $\pm$  4.5 &    -1.10 & -12.47 &  3.2 $\pm$ 0.95 & 21 &  1.31 $\pm$ 0.38 &  1.11 &    23.9 $\pm$     2.0 &   23.6 &  6.3 &  1.7 &    3 &  16.5 & 0.071$^{+0.083}_{-0.071 }$ \\
S999 &   24 $\pm$  2.7 &  -1.40 & -11.14 &  9.9 $\pm$  1.9 & 5.5 &   4.4 $\pm$ 0.86 &  1.03 &    23.3 $\pm$   1.3 &   20.7 &   13 &    1 &    2 &  16.5 &  1.1$^{+0.25}_{-0.24 }$ \\
VUCD6 &   21 $\pm$    3 &    -1.10 & -12.27 &    3.0 $\pm$  1.1 & 17 &  1.21 $\pm$ 0.45 &  1.18 &    25.2 $\pm$   1.8 &   16.4 &  2.9 &  2.1 &    3 &  16.5 & 0.039$^{+0.085}_{-0.039 }$ \\
S928 &   20 $\pm$  1.8 &  -1.30 & -11.64 &  5.3 $\pm$  1.3 & 8.8 &  2.34 $\pm$ 0.59 &  1.06 &    20.0 $\pm$   0.9 &   23.8 &   14 &  1.1 &    2 &  16.5 &  0.4$^{+0.15}_{-0.14 }$ \\
UCD5$^*$ &   20 $\pm$  4.5 &  -1.20 &  -12.10 &  3.4 $\pm$ 0.95 & 14 &  1.44 $\pm$ 0.41 &   1.10 &    22.9 $\pm$   2.6 &   26.2 & --- &  --- &  3.5 &    20 & 0.061$^{+0.054}_{-0.048 }$ \\
F1 &   18 $\pm$ 0.95 &  -0.64 & -12.29 &  2.5 $\pm$ 0.49 & 20 & 0.864 $\pm$ 0.17 &  1.05 &    22.2 $\pm$   0.6 &   24.2 &  2.3 &  2.3 &    8 &    20 &    0$^{+0.003}_{-   0 }$ \\
F9 &   17 $\pm$  1.9 &  -0.62 & -11.43 &  5.4 $\pm$    1.0 & 9.3 &  1.85 $\pm$ 0.36 &  1.05 &    28.4 $\pm$   1.6 &   9.53 &  3.3 &  1.5 &    8 &    20 & 0.21$^{+ 0.1}_{-0.085 }$ \\
S490 &   15 $\pm$  1.9 &  0.18 & -11.06 &  6.5 $\pm$  1.4 & 11 &   1.4 $\pm$ 0.29 &  1.02 &    42.1 $\pm$   2.7 &   3.74 &  0.7 &  1.8 &    2 &  16.5 & 0.088$^{+0.068}_{-0.058 }$ \\
F8$^*$ &   14 $\pm$  1.2 &  -0.35 &  -11.4 &  4.4 $\pm$  1.1 & 10 &  1.32 $\pm$ 0.32 &  1.23 &    30.2 $\pm$   1.3 &    6.7 & --- &  --- &    8 &    20 & 0.087$^{+0.085}_{-0.073 }$ \\
0330 &   12 $\pm$  2.2 &  -0.36 &    -11.00 &  5.7 $\pm$  1.3 & 7.1 &  1.72 $\pm$  0.4 & 0.859 &    41.5 $\pm$   3.7 &   3.26 & 0.86 &  1.7 &    7 &  3.77 & 0.17$^{+0.099}_{-0.09 }$ \\
UCD12a &  9.2 $\pm$  1.5 &  -0.35 &  -11.60 &  2.5 $\pm$ 0.71 & 12 & 0.739 $\pm$ 0.21 &   1.1 &    23.9 $\pm$   1.9 &   7.23 &  1.4 &  1.8 &    6 &    20 &    0$^{+0.015}_{-   0 }$ \\
F2$^*$ &  9.1 $\pm$  1.7 &  -0.73 & -11.45 &  2.8 $\pm$ 0.85 & 9 &  1.02 $\pm$ 0.31 &  1.12 &    19.9 $\pm$   1.8 &   14.5 & --- &  --- &    8 &    20 & 0.002$^{+0.035}_{-0.002 }$ \\
S314 &    9 $\pm$ 0.72 &  -0.50 & -10.97 &  4.3 $\pm$ 0.73 & 6.4 &   1.4 $\pm$ 0.24 & 0.987 &    35.2 $\pm$   1.4 &   3.32 & 0.84 &  1.7 &    2 &  16.5 &  0.1$^{+0.061}_{-0.057 }$ \\
F22 &  8.5 $\pm$ 0.87 &  -0.49 & -11.22 &  3.2 $\pm$ 0.91 & 8.1 &  1.04 $\pm$  0.3 &  1.14 &    29.1 $\pm$   1.5 &   5.34 & 0.49 &  2.3 &    8 &    20 & 0.007$^{+0.047}_{-0.007 }$ \\
0365 &  7.9 $\pm$  1.6 &  -0.90 & -11.05 &  3.5 $\pm$ 0.71 & 5.8 &  1.37 $\pm$ 0.39 & 0.827 &    23.7 $\pm$   2.4 &   7.37 &  1.4 &  1.8 &    7 &  3.77 & 0.069$^{+0.065}_{-0.053 }$ \\
F5 &  7.6 $\pm$  0.6 &  -0.34 & -11.83 &  1.7 $\pm$ 0.28 & 15 & 0.495 $\pm$ 0.083 &  1.06 &    25.4 $\pm$     1.0 &   5.24 &  1.3 &  1.8 &    8 &    20 &    0 \\
HGHH92-C1 &  6.8 $\pm$ 0.84 &  -1.2 &  -10.80 &  3.8 $\pm$ 0.96 & 4.2 &  1.62 $\pm$ 0.41 &     1.00 &    12.9 $\pm$   0.8 &   23.3 &    7 &  1.6 &    1 &  3.77 & 0.11$^{+0.078}_{-0.07 }$ \\
F17 &  6.2 $\pm$ 0.62 &  -0.55 & -11.37 &  2.1 $\pm$ 0.44 & 9.1 & 0.687 $\pm$ 0.15 &  1.03 &    28.0 $\pm$   1.4 &   3.46 & 0.94 &  1.7 &    8 &    20 &    0 \\
0265 &  5.4 $\pm$  1.5 &  -0.82 & -10.59 &  3.7 $\pm$  0.8 & 3.9 &  1.38 $\pm$ 0.49 & 0.953 &    20.8 $\pm$   2.9 &    5.6 &  2.2 &  1.4 &    7 &  3.77 & 0.096$^{+0.14}_{-0.096 }$ \\
H8005 &  5.2 $\pm$  2.2 &  -1.30 & -10.89 &  2.7 $\pm$  1.5 & 4.5 &  1.16 $\pm$ 0.66 &  1.09 &   9.1 $\pm$   1.9 &   29.5 &   14 &  1.3 &    2 &  16.5 & 0.043$^{+0.19}_{-0.043 }$ \\
0320 &  4.9 $\pm$ 0.69 &  -0.90 & -10.35 &  4.2 $\pm$  0.6 & 3.1 &  1.62 $\pm$ 0.23 & 0.785 &    20.0 $\pm$   1.4 &   6.84 &  1.2 &  1.9 &    7 &  3.77 & 0.13$^{+0.051}_{-0.05 }$ \\
F11 &  4.9 $\pm$ 0.41 &  -0.61 &  -11.60 &  1.3 $\pm$ 0.27 & 11 & 0.447 $\pm$ 0.092 &  1.05 &    23.7 $\pm$     1.0 &   3.77 &    1 &  1.7 &    8 &    20 &    0 \\
VHH81-C5 &  4.6 $\pm$  1.3 &  -1.40 & -10.54 &  3.3 $\pm$ 0.73 & 3.1 &  1.48 $\pm$ 0.53 & 0.925 &    15.8 $\pm$   2.2 &    8.8 &  5.9 & 0.89 &    1 &  3.77 & 0.12$^{+0.16}_{-0.12 }$ \\
F7 &    4 $\pm$    1 &  -1.20 & -11.23 &  1.5 $\pm$ 0.55 & 6.2 & 0.635 $\pm$ 0.24 &  1.05 &    12.2 $\pm$   1.6 &   11.9 &  7.2 &  1.2 &    8 &    20 &    0 \\
0077 &  3.8 $\pm$ 0.69 &  -0.36 & -10.34 &  3.3 $\pm$ 0.68 & 3.9 & 0.992 $\pm$  0.2 & 0.753 &    16.7 $\pm$   1.5 &   7.67 &  1.3 &  1.9 &    7 &  3.77 &    0$^{+0.039}_{-   0 }$ \\
0041 &  3.8 $\pm$ 0.78 &  -0.34 & -10.13 &    4.0 $\pm$ 0.87 & 3.2 &  1.18 $\pm$ 0.26 &  0.83 &    17.6 $\pm$   1.8 &   6.76 &  1.2 &  1.9 &    7 &  3.77 & 0.035$^{+0.056}_{-0.035 }$ \\
F51 &  3.7 $\pm$ 0.56 &  -0.23 & -10.66 &  2.3 $\pm$ 0.72 & 5.6 & 0.661 $\pm$  0.2 &  1.05 &    20.9 $\pm$   1.6 &    4.4 & 0.51 &  2.5 &    8 &    20 &    0 \\
F6 &  3.4 $\pm$    1 &  -1.30 & -11.17 &  1.3 $\pm$ 0.51 & 5.7 & 0.591 $\pm$ 0.22 &  1.05 &    14.0 $\pm$   2.1 &   7.64 &  2.4 &  1.6 &    8 &    20 &    0 \\
HGHH92-C6 &  3.3 $\pm$ 0.48 &  -0.91 & -11.01 &  1.5 $\pm$ 0.37 & 5.6 & 0.596 $\pm$ 0.14 & 0.929 &    20.7 $\pm$   1.5 &    4.3 & 0.57 &  2.4 &    1 &  3.77 &    0 \\
0326 &  3.1 $\pm$ 0.57 &    -1.0 & -10.07 &  3.4 $\pm$ 0.69 & 2.3 &  1.35 $\pm$ 0.27 & 0.876 &    19.4 $\pm$   1.8 &   3.73 &  1.1 &  1.6 &    7 &  3.77 & 0.082$^{+0.067}_{-0.062 }$ \\
F34 &  2.2 $\pm$ 0.43 &  -0.77 & -10.83 &  1.2 $\pm$ 0.37 & 5 & 0.441 $\pm$ 0.13 &  1.06 &    15.3 $\pm$   1.5 &   4.19 &  1.4 &  1.5 &    8 &    20 &    0 \\
F53 &  2.1 $\pm$ 0.53 &  -0.80 & -10.65 &  1.4 $\pm$ 0.49 & 4.2 & 0.513 $\pm$ 0.18 &  1.05 &    14.5 $\pm$   1.8 &   4.71 & 0.95 &    2 &    8 &    20 &    0 \\
VHH81-C3 &  1.9 $\pm$ 0.27 &  -0.22 & -10.51 &  1.4 $\pm$ 0.37 & 4.9 & 0.397 $\pm$  0.1 & 0.815 &    16.1 $\pm$   1.1 &    4.2 & 0.47 &  2.6 &    1 &  3.77 &    0 \\
0372 &  1.9 $\pm$ 0.42 &  -1.20 &  -9.84 &  2.6 $\pm$ 0.73 & 1.7 &  1.12 $\pm$ 0.31 & 0.769 &    16.5 $\pm$   1.8 &   3.54 & 0.68 &    2 &    7 &  3.77 & 0.023$^{+0.068}_{-0.023 }$ \\
0227 &  1.9 $\pm$ 0.33 &  -1.30 &  -9.56 &  3.3 $\pm$  0.6 & 1.3 &  1.46 $\pm$ 0.26 & 0.784 &    13.6 $\pm$   1.2 &    5.6 &  1.2 &  1.7 &    7 &  3.77 & 0.099$^{+0.075}_{-0.057 }$ \\
0218 &  1.7 $\pm$ 0.43 &  -0.49 &  -9.69 &  2.7 $\pm$ 0.88 & 2 & 0.863 $\pm$ 0.28 & 0.817 &    12.7 $\pm$   1.6 &   5.23 &  1.2 &  1.9 &    7 &  3.77 &    0$^{+0.032}_{-   0 }$ \\
0040 &  1.6 $\pm$  0.3 &  -0.41 &  -9.69 &  2.4 $\pm$ 0.54 & 2.1 &  0.76 $\pm$ 0.17 & 0.825 &    13.7 $\pm$   1.3 &   4.43 & 0.77 &  1.9 &    7 &  3.77 &    0 \\
0378 &  1.3 $\pm$ 0.25 &  -0.51 &  -9.79 &  1.9 $\pm$ 0.56 & 2.2 & 0.623 $\pm$ 0.18 & 0.895 &    14.2 $\pm$   1.3 &   3.26 & 0.58 &  1.9 &    7 &  3.77 &    0 \\
0150 &  1.1 $\pm$ 0.17 &  -0.54 & -10.05 &  1.2 $\pm$  0.2 & 2.7 & 0.409 $\pm$ 0.067 &  0.85 &    15.9 $\pm$   1.2 &   1.92 &  0.4 &    2 &    7 &  3.77 &    0 \\
0232 &  1.1 $\pm$ 0.31 &  -0.77 &  -9.45 &    2 $\pm$ 0.78 & 1.4 & 0.753 $\pm$ 0.28 & 0.812 &    14.1 $\pm$   2.1 &    2.6 &  0.5 &    2 &    7 &  3.77 &    0$^{+0.011}_{-   0 }$ \\

\normalsize
\\\hline
\end{longtable}
\end{landscape}
\end{longtab}

\begin{figure*}[]
\begin{center}
\includegraphics[width=6.1cm]{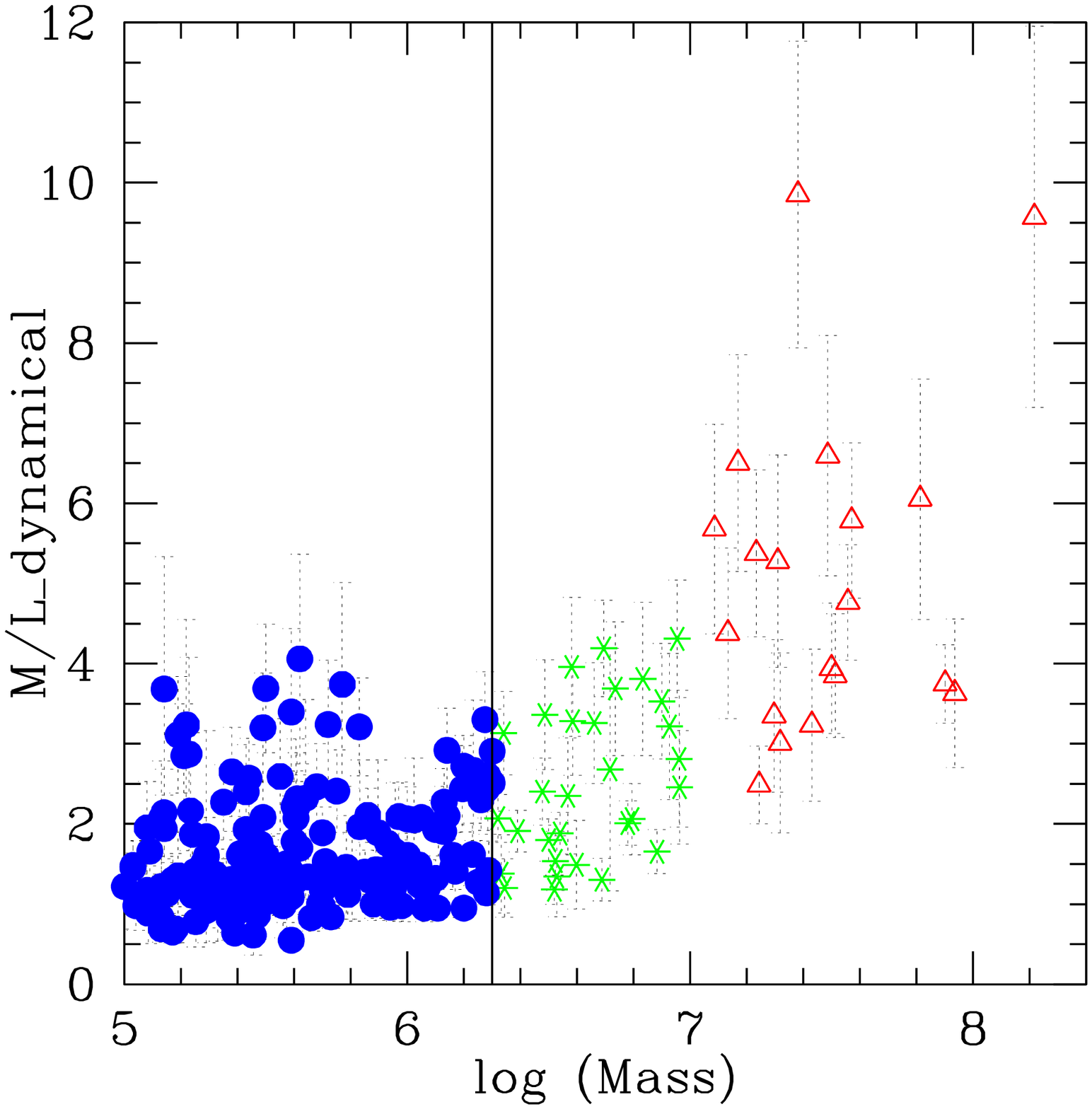}
\includegraphics[width=6.1cm]{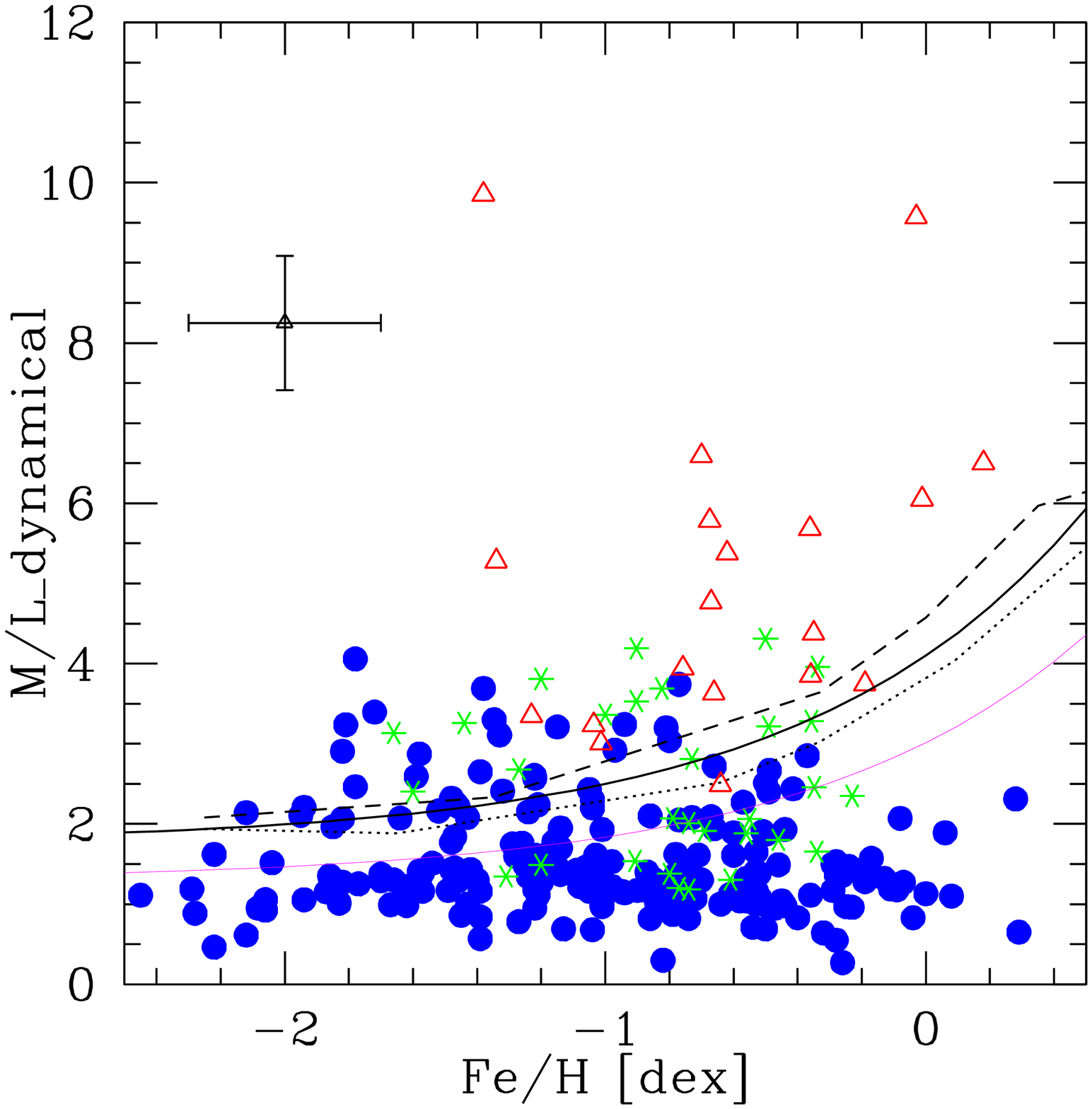}
\includegraphics[width=6.1cm]{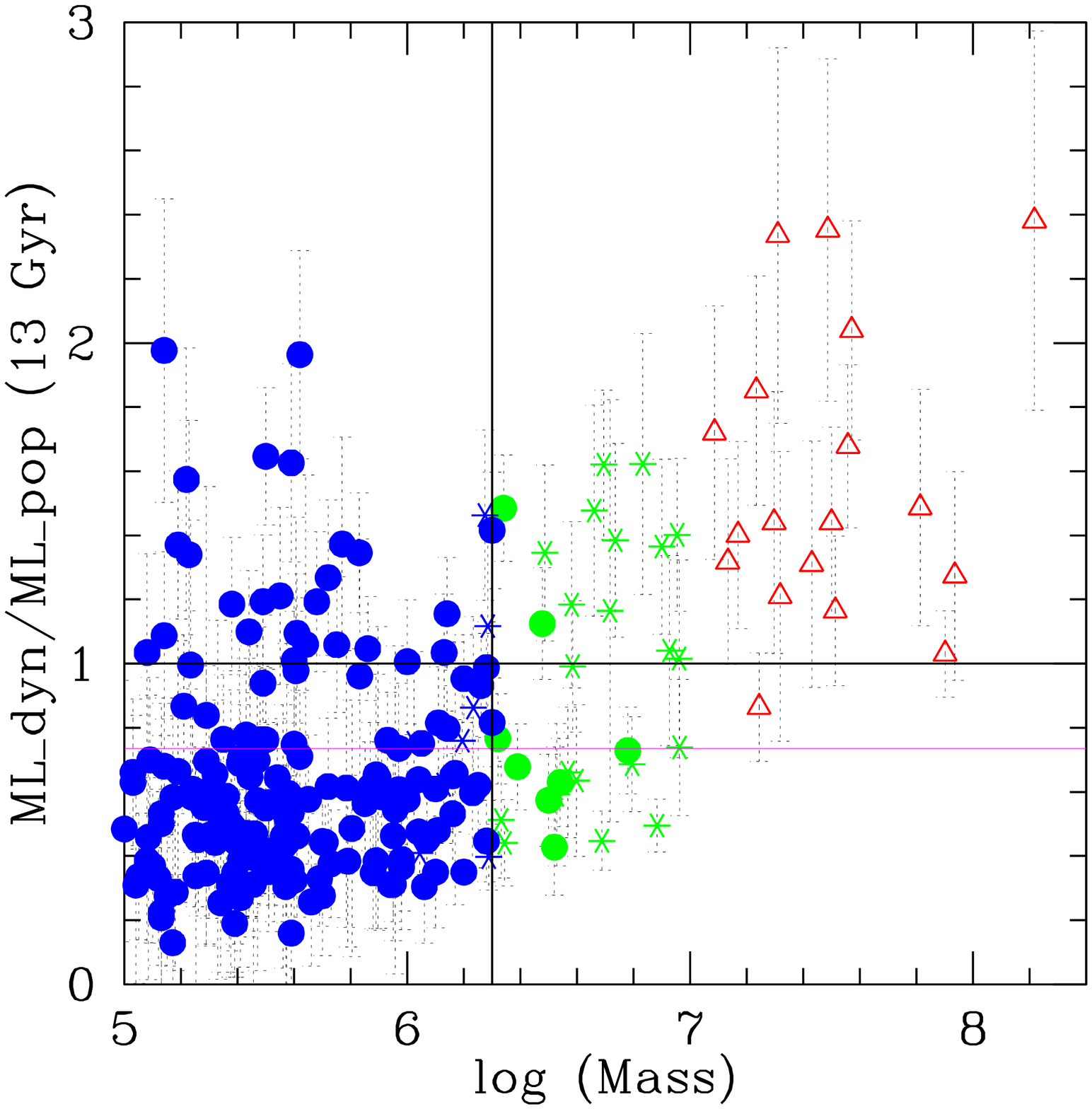}
\caption{{\bf Left panel:} Dynamical mass-to-light ratios of compact stellar systems plotted vs. their dynamical mass. Data taken from the literature as indicated in the text. Sources are colour coded according to mass. GCs ($M<2*10^6 \msun$) are blue, low-mass UCDs ($2*10^6 \msun < M <10^7 \msun$) are green, high-mass UCDs ($ M > 10^7 \msun$) are red. GC data points are dominated by Local Group measurements, whereas UCD data points are mainly extragalactic (Fornax/Virgo/CenA). {\bf Middle panel}: Dynamical mass-to-light ratios of compact stellar systems plotted vs. their metallicities. Sample and symbols as in the left panel. The dotted lines correspond to predicted stellar M/L ratios for a 13 Gyr stellar population of solar $\alpha$ abundance; top line is from Maraston et al. (\citeyear{Marast05}), bottom line is from Bruzual \& Charlot (\citeyear{Bruzua03}). The solid curve in between is an exponential fit to the mean of the two dotted lines. It represents the maximum stellar M/L for a canonical IMF (Kroupa/Chabrier). Sources with dynamical M/L ratios above that solid line need some extra dark mass, and/or IMF variation. The (magenta) thin solid line below the thick solid line corresponds to the stellar M/L ratio for a younger, 9 Gyr stellar population of solar $\alpha$ abundance. {\bf Right panel:} The dimensionless quantity $\Psi=(M/L)_{\rm dyn}/(M/L)_{\rm pop}$ is plotted vs. dynamical mass for the same sources as in previous plots, using equation~\ref{mlstellar} for $(M/L)_{\rm pop}$, see also middle panel. The black horizontal line indicates $\Psi=1$ for 13 Gyr assumed age (see also black solid line in middle panel), the (magenta) thin solid line indicates $\Psi=1$ for 9 Gyr assumed age (magenta solid line in middle panel). We mark UCDs ($M>2*10^6 \msun$) from the Local Group with (green) filled circles, and massive extragalactic GCs ($M<2*10^6 \msun$) with (blue) asterisks.}
\label{massML}
\end{center}
\end{figure*}

\subsection{Distribution of dynamical M/L vs. mass}
\label{distribution}
In Table~\ref{tableall} we list the updated dynamical masses and
dynamical M/L ratios for the compact stellar systems considered, along
with the mass difference to the original publication. We also report
their absolute magnitudes, stellar population masses, literature measured
velocity dispersion, effective radius, metallicity, adopted structural
profile parameters and distance used in the modelling.

In the left panel of Fig.~\ref{massML} we plot the updated
measurements of mass versus dynamical M/L ratio of the above
sources. For comparison we also include the most recent data on
compact stellar systems for the Milky Way (Mc Laughlin \& van der
Marel~\citeyear{McLaug05}) and M31 (Strader et
al.~\citeyear{Strade09}). We reproduce the previously known trend
(Mieske et al.~\citeyear{Mieske08}) that the dynamical M/L of compact
stellar systems above $2\times 10^6 \msun$ - the regime of UCDs -
increase with mass. We colour code the sources according to mass: GCs
($M<2*10^6 \msun$) are blue, low-mass UCDs ($2 \times 10^6 \msun < M <10^7
\msun$) are green, high-mass UCDs ($ M > 10^7 \msun$) are red.

In the middle panel of Fig.~\ref{massML} we plot the dynamical M/L
ratio vs. metallicity [Fe/H]. The lines correspond to predicted
stellar M/L ratios for a 13 Gyr stellar population of solar $\alpha$
abundance; dashed line is from Maraston et al. (\citeyear{Marast05}),
dotted line is from Bruzual \& Charlot (\citeyear{Bruzua03}). The
solid line is the mean of exponential fits to the two models, given
by the quantity $(M/L)_{\rm pop}$ introduced in equation~\ref{mlstellar},
thus being the maximum stellar M/L for a canonical
(e.g. Kroupa~\citeyear{Kroupa02}) IMF.

It is evident that many GCs are located significantly below that ridge
line, in particular the more metal-rich GCs. This was reported already
by Strader et al. (\citeyear{Strade09}, \citeyear{Strade11}). This is
interesting in its own right and may indicate an IMF variation. In the
present paper, we focus however on the most massive compact stellar
systems, the UCDs. Almost all of the massive UCDs have dynamical M/L
above the ridge line, consistent with dark mass on top of a canonical
IMF.

In the right panel of Fig.~\ref{massML} we plot for each source the
dimensionless quantity
\begin{equation}
\Psi=(M/L)_{\rm dyn}/(M/L)_{\rm pop}
\end{equation}
defined as the ratio between dynamical and stellar population M/L
ratio. $\Psi$ above unity indicates the presence of non-luminous mass
on top of a canonical stellar population (assuming virial
equilibrium). For calculating $\Psi$, the dynamical M/L ratio is taken
from the top left panel, and the stellar population M/L ratio is taken
from equation~\ref{mlstellar}.

The plotted uncertainty of $\Psi$ is a Gaussian propagation of two
main error contributions. 1. the uncertainty of the dynamical M/L,
which itself encompasses the error bar of the velocity dispersion,
distance estimate, luminosity, surface brightness profile. 2. An
assumed global 0.3 dex uncertainty in the sources' [Fe/H], translated
to an uncertainty factor of $(M/L)_{\rm pop}$ (Mieske et
al.~\citeyear{Mieske08}).
\begin{figure}[]
\begin{center}
\includegraphics[width=8.6cm]{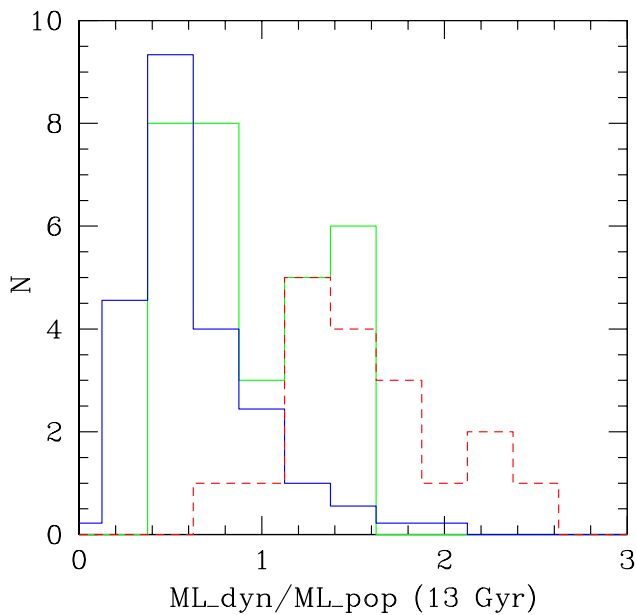}
  \caption{Histogram of $\Psi=(M/L)_{\rm dyn}/(M/L)_{\rm pop}$ for the samples of Fig.~\ref{massML}. The (dashed) red histogram denotes high-mass UCDs with $M>10^7 \msun$. The green (solid) histogram denotes low-mass UCDs $2 \times 10^6 \msun < M <10^7 \msun$. The blue (solid) hostogram denotes globular clusters ($M<2 \times 10^6 \msun$). The histogram for GCs is scaled down by a factor of 9 to fit the plot. The low $\Psi$ peak of the green histogram is close to the peak of the blue (GC) sample. The shift of $\sim$20\% between those two peaks is expected from dynamical evolution (e.g. Baumgardt \& Makino~\citeyear{Baumga03}, Kruijssen \& Mieske~\citeyear{Kruijs10}).}
\label{MLhist}
\end{center}
\end{figure}

\subsubsection{A bimodal M/L distribution for low-mass UCDs}
\label{bimodal}
In Fig.~\ref{MLhist} we show the histogram of $\Psi$ for GCs, low-mass
UCDs, and high-mass UCDs. The 19 high-mass UCDs ($ M > 10^7 \msun$) have a
median $\Psi$ of 1.45, with a mean of 1.73 $\pm$ 0.19. The 30 low-mass
UCDs ($ M < 10^7 \msun$) have an average $\Psi$ of 0.93 $\pm 0.07$,
consistent with, on average, a 'normal' stellar population. From the
histogram morphology there is a hint for a bimodal distribution in
$\Psi$ in the low-mass sample with a dip around $\Psi \sim 1$. We
  use the KMM tool (e.g. Ashman et al.~\citeyear{Ashman94}) to test the
  significance of this possible bimodality in the $\Psi$
  distribution. We run KMM in homoscedastic mode, which means that a
  single (Gaussian) width is assumed for each of the two potential
  subpopulations. The KMM test yields the following results: the
  low-mass sample is inconsistent with a single Gaussian distribution
  in $\Psi$ at the 99.2\% confidence level. KMM finds two distinct
  Gaussian peaks at $\Psi=0.63$ and $\Psi=1.33$ of width
  $\sigma=0.17$, containing 14 and 16 UCDs, respectively. In contrast,
  the high-mass sample cannot be fitted with a bimodal Gaussian. All
  19 sources in this sample are assigned to one single Gaussian by
  KMM, peaked at $\Psi=1.73$ with a width $\sigma=0.77$.

The bimodal $\Psi$ distribution for low-mass UCDs and the offset
unimodal $\Psi$ distribution for high-mass UCDs are an intriguing
finding. It is consistent with the hypothesis that in the low-mass UCD
regime there is indeed an overlap between massive globular clusters
(with normal M/L ratios) and UCDs as stripped nuclei (with elevated
M/L ratios), and that the latter population dominates the high-mass
UCDs.

\subsubsection{On the assumption of single stellar populations}
\label{singlepop}
There is one potential concern when comparing the observed dynamical
M/L of UCDs to single stellar population (SSP) predictions of stellar
M/L: if UCDs are remnant nuclei of galaxies, or resemble massive GCs,
it is likely that they contain multiple stellar populations. Is it
therefore appropriate to compare the dynamical M/L to the predictions
of SSP models?

If a young population is present on top of an underlying old
population, the integrated light in the optical, and therefore the
spectroscopic population parameters available in the literature, will
be dominated by the young population, even if this population
contributes only a small fraction of the stellar mass. Thus, based on
spectroscopic population parameters, the mass-weighted age of
composite population UCDs would be underestimated. Assuming a
reasonable enrichment history with younger populations being at least
as enriched as older populations, their mass-weighted metallicities
would be overestimated by the spectroscopic value. Moreover, also
purely old, metal-poor ([Fe/H]$ \lesssim -1.2$ dex) stellar
populations can mimic a young population in integrated-light
spectroscopy, if they contain blue horizontal branch stars
(Ocvirk~\citeyear{Ocvirk10}). Since younger and/or more metal-poor
populations have a lower stellar M/L, underestimating the metallicity
or age would lead to an underestimation of the true mass-weighted
stellar population M/L$_{\rm pop}$.

However, if one uses only spectroscopic
metallicity measurements (that will tend to be over- but \emph{not}
underestimated in the presence of younger stellar populations), and
adopts an old age for safety, even if spectroscopic indices may
indicate differently, one will not underestimate, but potentially
overestimate, the M/L$_{\rm pop}$ of UCDs. 

Hence, the apparent surplus of dynamical M/L for massive UCDs cannot
be explained as the result of the presence of multiple stellar
populations. In fact, in the presence of multiple populations, the
discrepancy between dynamical and stellar population M/L may be even
higher than our estimates.

\begin{figure*}[]
\begin{center}
\includegraphics[width=5.8cm]{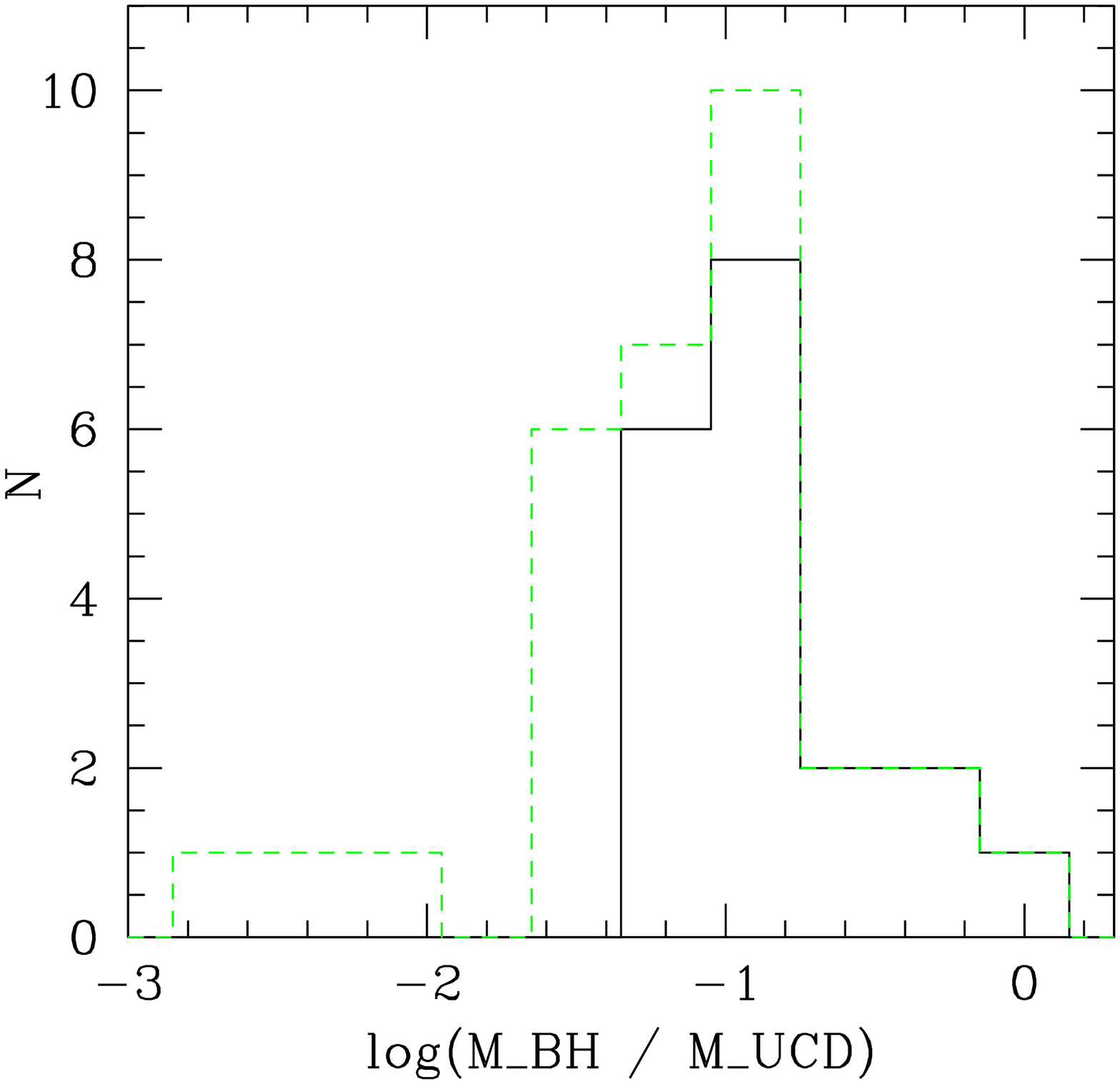}
\includegraphics[width=5.8cm]{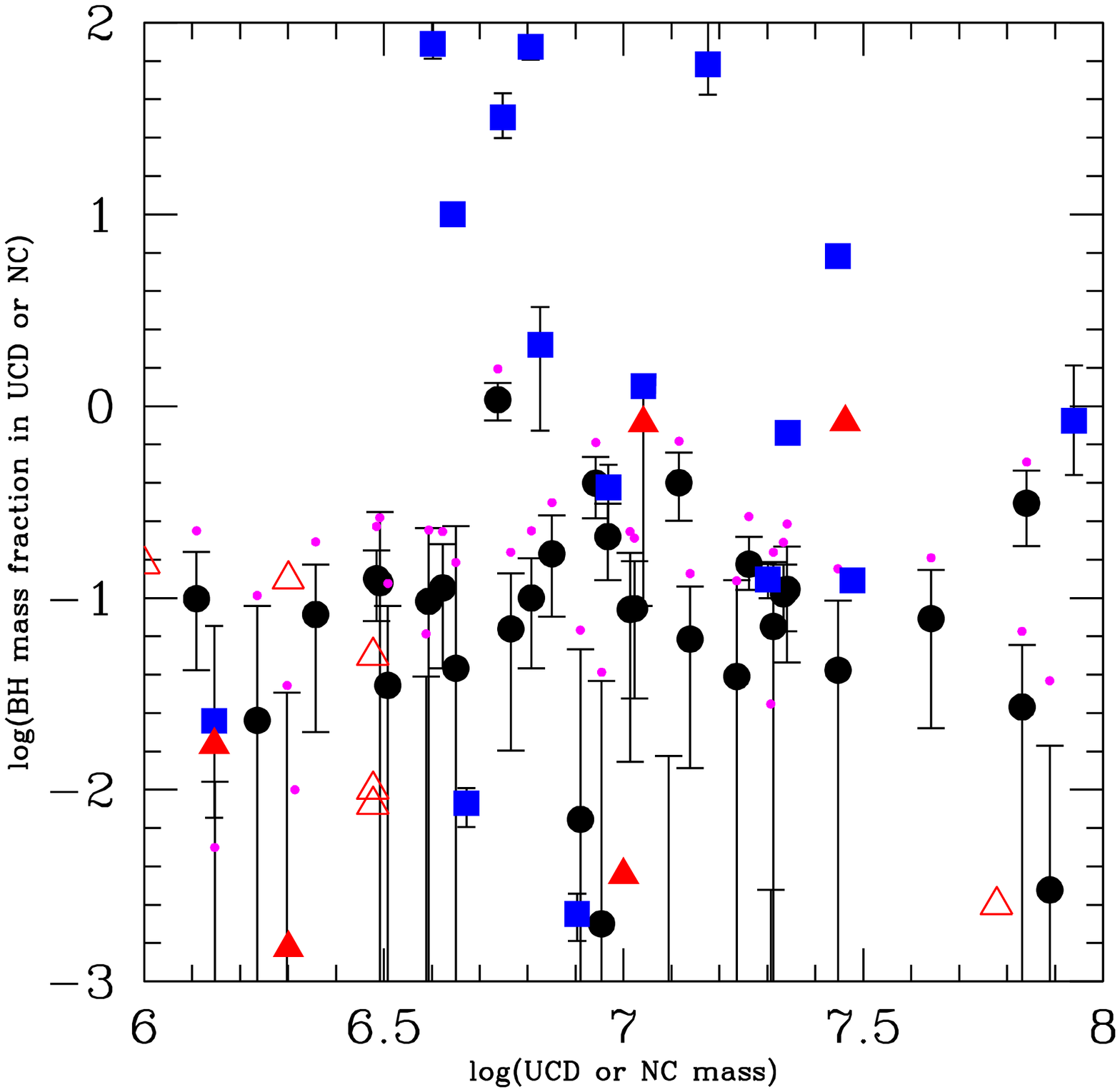}
\includegraphics[width=5.8cm]{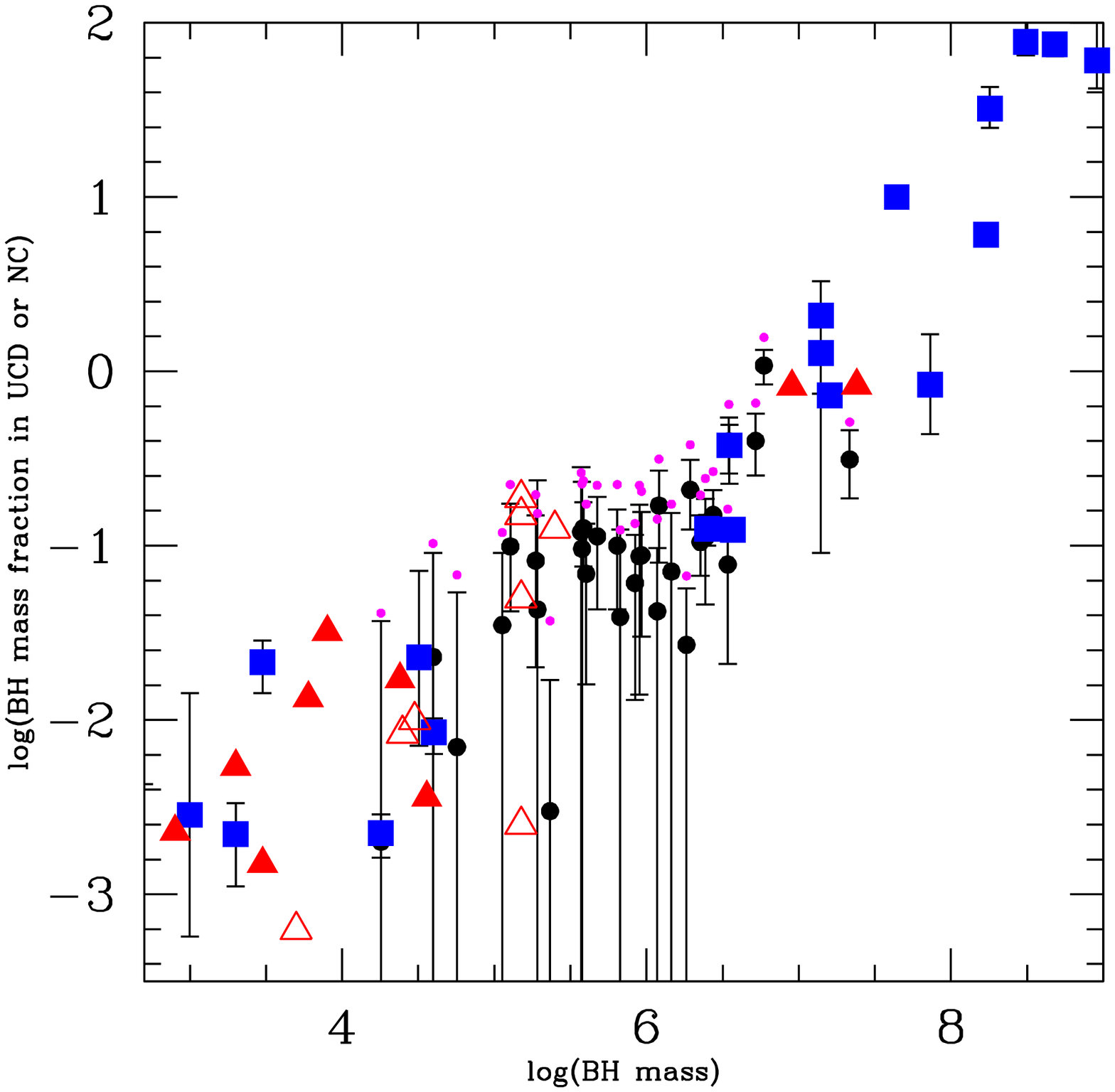}\\

  \caption{{\bf Left panel:} Histogram of the black hole mass fraction
    as estimated for those UCDs with $\Psi=(M/L)_{\rm dyn}/(M/L)_{\rm pop}>1$,
    see Sect.~\ref{BH}. The solid histogram corresponds to BH mass
    estimates whose lower 1 $\sigma$ error bound is above 0. The
    dashed (green) histogram includes also those UCDs for which the
    lower 1 $\sigma$ error bound is consistent with 0.  {\bf Middle
      panel:} Black hole mass fraction in compact stellar systems
    plotted vs. the stellar mass of the compact stellar
    systems. Filled black circles are the UCDs from the left
    panel. The UCD mass used on the x-axis is the stellar population
    mass M$_{\rm pop}$ for an age of 13 Gyr. The small filled (magenta)
    circles indicate the modelled BH mass assuming a stellar
    population age of 9 Gyr. Filled blue squares are BH mass
    measurements in nuclear clusters from Graham et
    al. (\citeyear{Graham09}) which have a lower BH mass limit larger
    than 0. Filled (red) triangles are those sources from Graham et
    al. for which the BH mass is an upper limit. Open (red) triangles
    are upper limits for BHs in lower mass nuclear clusters from
    Neumayer \& Walcher (\citeyear{Neumay11}). {\bf Right panel:}
    Black hole mass fraction in compact stellar systems plotted
    vs. the black hole mass, for the same objects as in the middle
    panel.}
\label{BHrelmass}
\end{center}
\end{figure*}

\begin{figure*}[]
\begin{center}
\includegraphics[width=8.6cm]{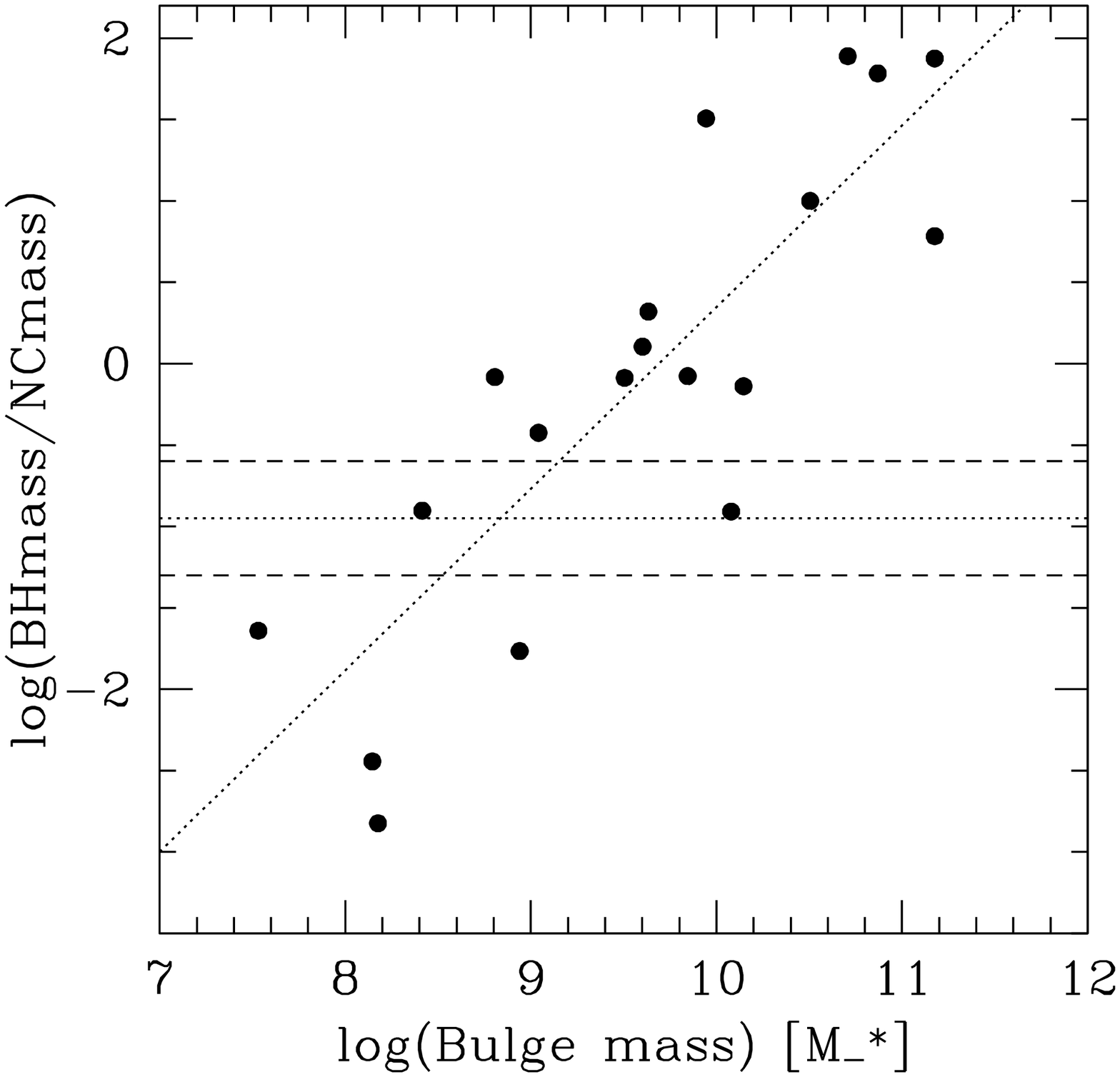}
\includegraphics[width=8.6cm]{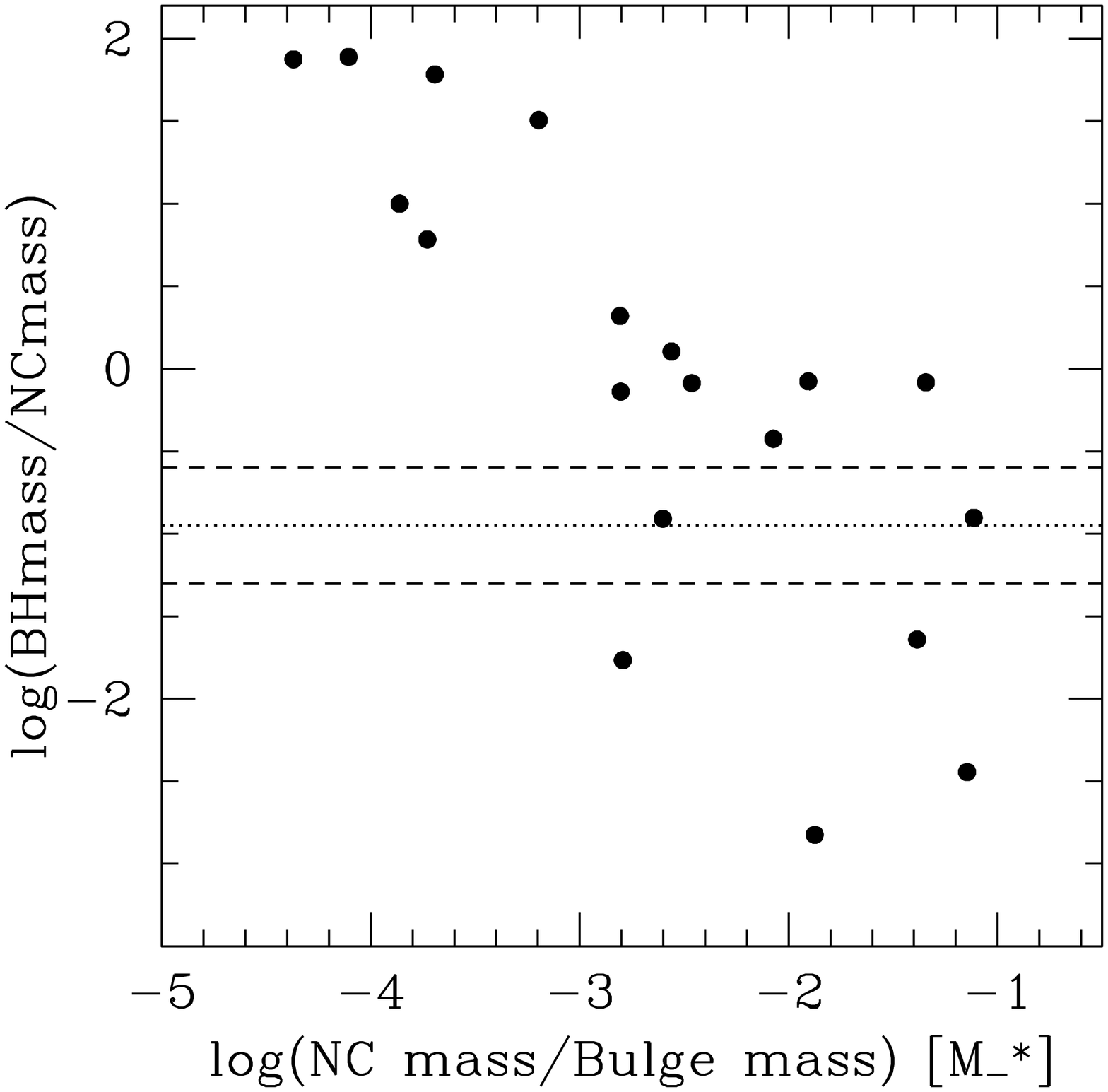}\\
  \caption{{\bf Left panel:} For those galaxies with a BH and a NC in the sample of Graham et al. (\citeyear{Graham09}), we plot the ratio between BH and NC mass vs. the total bulge mass of the host galaxy. A least square linear fit through the data points is indicated as dotted line. The horizontal dotted and dashed lines indicate the median and rms scatter of BH mass fraction in UCDs, see Fig.~\ref{BHrel_King}. The typical host galaxy bulge masses for those nuclei that have a BH mass fraction comparable to UCDs is 10$^{9 \pm 1}$ solar masses. {\bf Right panel:} Analogous to the left plot, we give on the x-axis the ratio of NC over host galaxy bulge mass. The typical NC/bulge mass fraction is $\sim$1\% in the range of BH/NC mass fraction of UCDs. We do not plot a linear fit to the literature data due to their larger scatter compared to the left panel.}
\label{NCrelmass}
\end{center}
\end{figure*}

\begin{figure*}[]
\includegraphics[width=8.6cm]{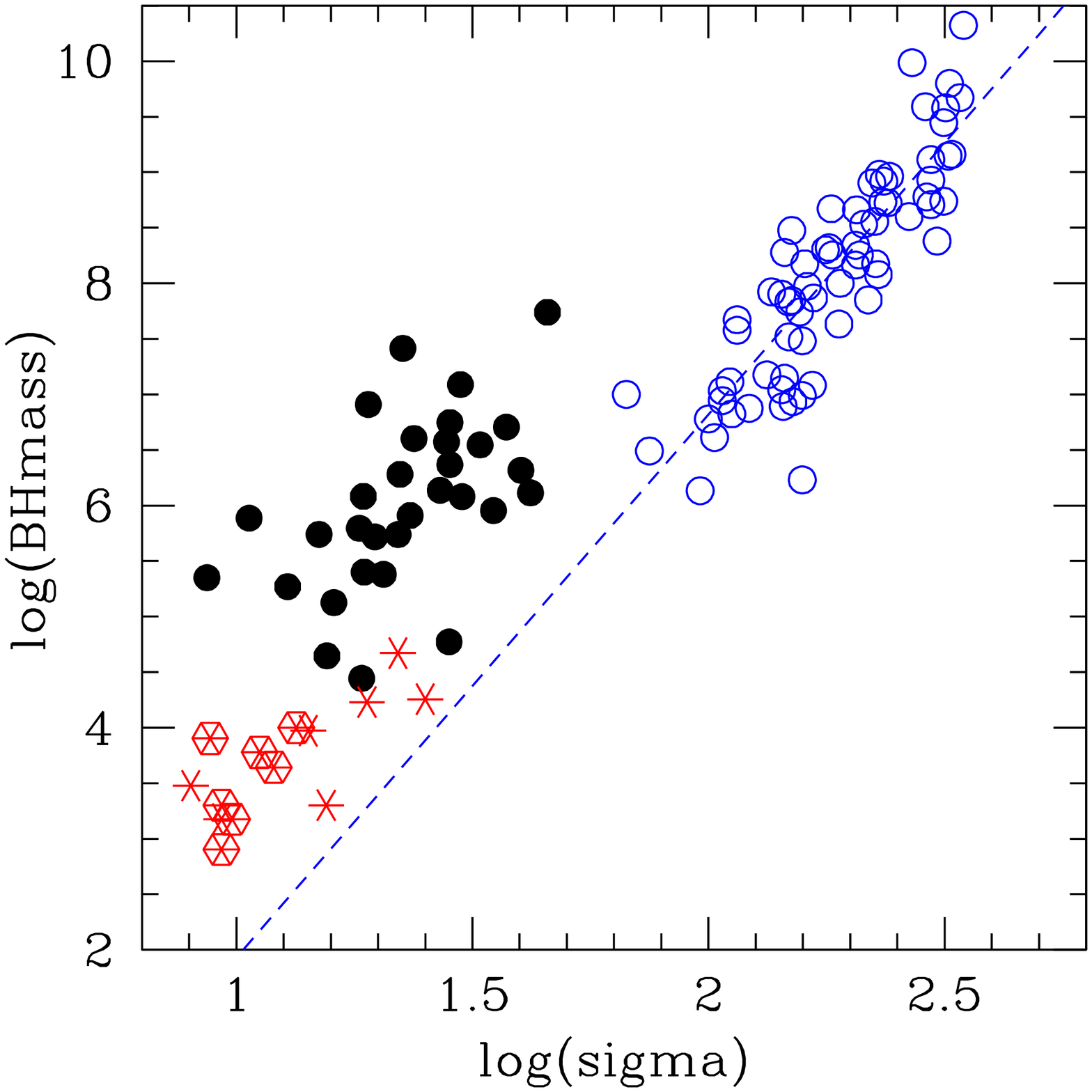}
\includegraphics[width=8.6cm]{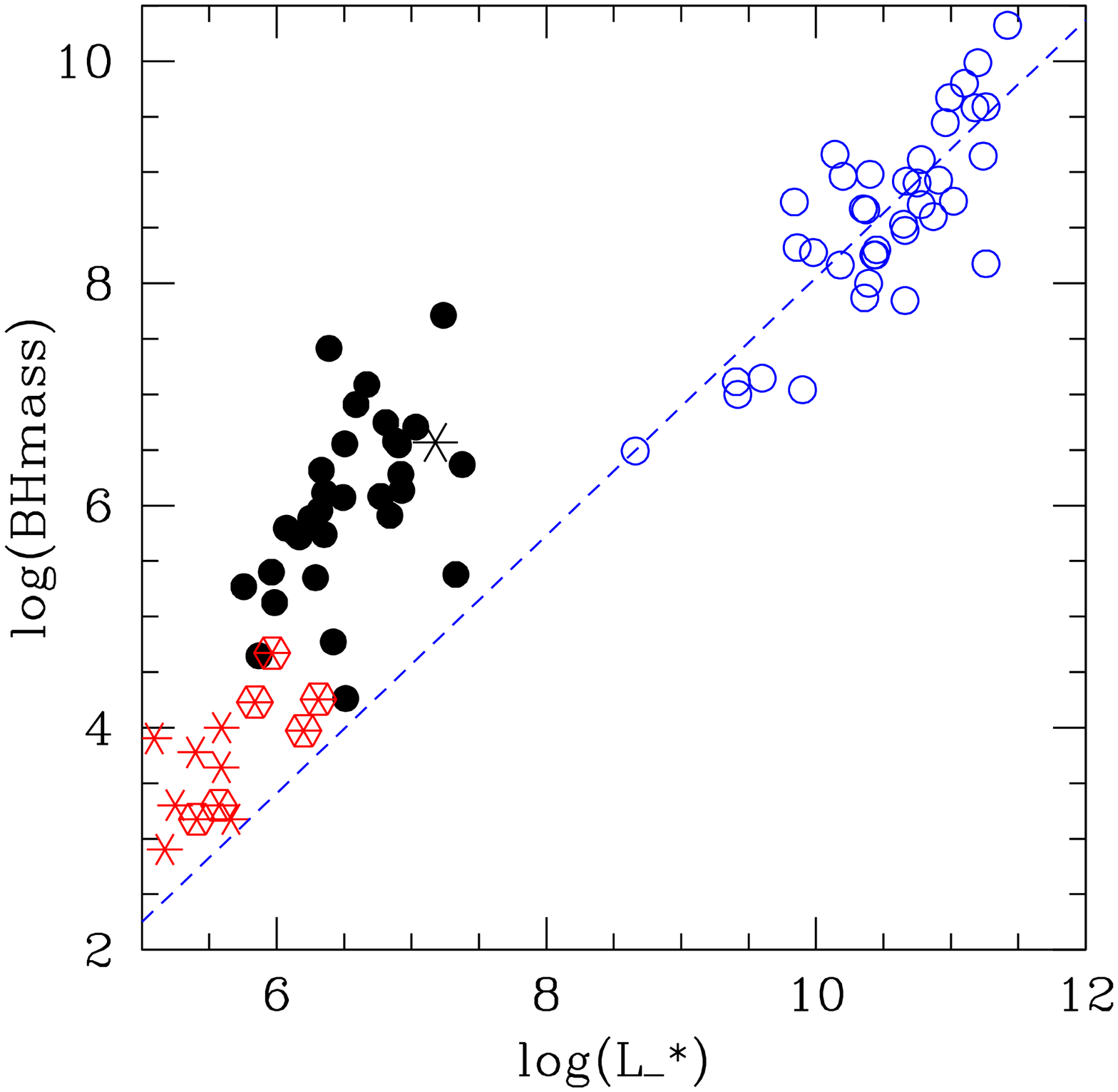}
\caption{{\bf Left Panel:} M$_{\rm BH}$-$\sigma$ relation for UCDs
  (black filled circles), compared to Galactic GCs
  (red asterisks, L\"{u}tzgendorf et al.~\citeyear{Luetzg13}; sources
  with hexagons have only upper limits to their BH mass) and galaxies
  (blue circles; McConnell et al.~\citeyear{McConn13}). The blue
  dashed line is a linear fit in log-log space to the galaxy
  sample. It has a slope of 4.9. The BH mass of UCDs is on average
  offset by about 2 dex from this relation. {\bf Right panel:} M$_{\rm
    BH}$-Luminosity relation for the same sources as in the left
  panel. Some objects from the galaxy sample of McConnell et al. are
  not included here since they lack information on host galaxy
  luminosity. The blue dashed line is a linear fit in log-log space to
  the galaxy sample with a slope of 1.16, indicating near direct
  proportionality between BH mass and host luminosity. The luminosity
  of UCDs is on average offset by 2 dex with respect to this
  reference. For illustration purposes we also show as large asterisk
  the central black hole mass of the Milky Way vs. the luminosity of
  the Milky Way's nuclear cluster (Graham \&
  Spitler~\citeyear{Graham09}).}
\label{sigmass}
\end{figure*}

\section{BH mass estimates in UCDs}
\label{BH}
In the following we provide an estimate of the mass of a hypothetical
central BH for those UCDs whose dynamical mass exceeds the
stellar population mass, that is for those UCDs with $\Psi=(M/L)_{\rm
  dyn}/(M/L)_{\rm pop} > 1$ in Fig.~\ref{massML} right panel. In doing
so, we assume that the gravitational potential of the BH is
responsible for the elevated overall velocity dispersion which gives
rise to the nominally high dynamical M/L:

\begin{enumerate}    

\item We continue after the first step outlined in
  Sect.~\ref{procedure} for determining the observed velocity
  dispersion $\sigma_{\rm \rm obs,pop}$ for the stellar population mass of the
  UCDs.

\item In order to determine BH masses for the UCDs, we add point-mass
  potentials $-G M_{\rm BH}/r$ with varying mass fractions $M_{\rm BH}/M_{\rm UCD}$ to the
  potentials based on the stellar distribution calculated in step 2 of
  sec. 2.2 of each UCD. We then determine new distribution functions
  f(E) and velocity distributions for the stars in the combined
  potentials.

\item We then calculate the predicted velocity dispersions seen by the
  observers as outlined in steps 3 to 5 of sec. 2.2 and determine the
  BH mass fractions which lead to an agreement of the observed
  velocity dispersion with the predicted ones.

\item To estimate the uncertainty of the BH mass, we assign the full
  measurement uncertainty of the dynamical mass-to-light ratio to the
  observed velocity dispersion, which can be denoted as $\Delta
  \sigma_{\rm obs,total}$, and redetermine the BH mass for $\sigma_{\rm \rm
    obs,new} = \sigma_{\rm \rm obs} \pm \Delta \sigma_{\rm obs,total}$. To
  obtain $\Delta \sigma_{\rm obs,total}$, we first separate the literature
  uncertainty $\Delta \sigma_{\rm obs}$ of the observed velocity
  dispersion $\sigma_{\rm obs}$ from the overall error of
  $\Psi$. For this we assume Gaussian propagation, and
  that the relative contribution of the $\sigma$ error to the M/L
  uncertainty is $2 \frac{\Delta \sigma}{\sigma}$. The resulting
  residual error is the uncertainty of the stellar population mass, and we
  denote it $\Delta_{\rm Mpop}$. For a constant BH mass fraction, an
  uncertainty $\Delta_{\rm Mpop}$ in the stellar population mass will lead to a
  change of the predicted velocity dispersion by $\Delta
  \sigma_{\rm obs,from M/L} = \sigma
  (\sqrt{1+\frac{\Delta_{\rm Mpop}}{M_{pop}}}-1)$. Since this
  uncertainty is statistically independent of the measurement
  uncertainty $\Delta \sigma_{\rm obs}$, it will lead to a combined
  uncertainty $\Delta \sigma_{\rm obs,total} = \sqrt{\Delta
    \sigma_{\rm obs,from M/L}^2+\Delta \sigma_{\rm obs}^2}$.

\item Not all of the studied UCDs may necessarily be consistent with
  an assumed age of 13 Gyrs, which is a conservative assumption in the
  context of this study - $\Psi$ is lower for older ages. We
  therefore repeat the above steps also for a stellar population mass
  corresponding to a younger age of 9 Gyr. According to the models of
  Bruzual \& Charlot~(\citeyear{Bruzua03}) and
  Maraston~(\citeyear{Marast05}) the expected stellar population mass is a
  factor of 1.36 below the stellar population mass for 13 Gyr, see
  Fig.~\ref{massML}. That is, the BH mass obtained for an assumed age
  of 9 Gyr will always be above the BH mass for 13 Gyr, since
  $\Psi_{\rm 9Gyr}=1.36 \times \Psi_{\rm 13Gyr}$.

\end{enumerate}

\noindent The resulting histogram of inferred BH mass fractions is shown in the
left panel of Fig.~\ref{BHrelmass}. The median BH mass fraction for
UCDs with significant BH masses (lower error bar above 0) is 11\%. In
the left and middle panels of that figure we compare these fractions
to literature data of BH masses in galaxy nuclear clusters (NCs;
Graham et. al.~\citeyear{Graham09}, Neumayer \&
Walcher~\citeyear{Neumay11}). We note that the BH masses from Neumayer
\& Walcher are all upper limits. The middle panel shows the BH mass
fraction vs. UCD and NC mass. The right panel shows the BH mass
fraction vs. the BH mass. In those two panels we also show as small
(magenta) dots the BH mass estimates for UCDs assuming an age of 9 Gyr
for the stellar population. The median BH mass for this case is 14\%.

\section{Discussion}
\label{discussion}

\subsection{BH mass fractions in UCDs and NCs, and the link to host galaxies}
One commonly discussed formation scenario of UCDs is the tidal
stripping of a more massive host galaxy, which leaves only the most
compact central object (its nucleus) intact (e.g.Bekki et
al.~\citeyear{Bekki03}, Pfeffer \&
Baumgardt~\citeyear{Pfeffe13}). Graham et al. (\citeyear{Graham09})
provide a compilation of literature measurements for galaxies that
have both a nuclear cluster and a central black hole.  In the left
panel of Fig.~\ref{NCrelmass} we plot the ratio of BH mass over NC
mass vs. the bulge mass of the respective host galaxies. We indicate as
dotted line a linear least square fit to the data. As horizontal
dashed lines we indicate the typical BH mass fraction that we
estimate for UCDs in the previous section.

If UCDs are 'naked nuclei' harbouring central BHs as fossil relics of
their progenitor galaxies, the BH masses can be used to constrain their
origin. With Fig.~\ref{NCrelmass} one can pose the question: what is
the host galaxy mass for NCs that contain a BH of about 10\% of the
NC's mass?  From the left panel of Fig.~\ref{NCrelmass} one thus
estimates, in the framework of the stripping hypothesis, a typical UCD
host mass of 10$^9 \msun$, with a range between 10$^8 \msun$ to
10$^{10} \msun$ solar masses. The right panel of the same figure shows
that this corresponds to a typical mass fraction of $\sim$1\% between
NC and host. Numerical simulations of galaxy tidal stripping
predict a mass fraction of a few percent between tidally stripped
naked nucleus and the original host (e.g. Bekki et
al.~\citeyear{Bekki03}, Pfeffer \& Baumgardt~\citeyear{Pfeffe13}),
which also leads to typical progenitor masses of $\sim 10^9 \msun$.

Thus, the BH mass fractions of around 10\% estimated for UCDs
based on their elevated M/L ratios agree well with BH mass fractions
of 'naked nuclei' expected in the tidal stripping scenario. At
  the same time, such BH mass fractions rule out the hypothesis that
  most UCDs are hyper-compact stellar systems bound to recoiling
  super-massive black holes that were ejected from the centres of
  massive galaxies (Merritt et al.~\citeyear{Merrit09}). For such
  hypothetical objects one would expect BH masses comparable to or
  larger than the stellar mass.

\subsection{Fraction of UCDs with significant BH masses}

From our full sample of 49 UCDs, we find 18 UCDs are consistent with a
formal BH mass of zero and 31 sources require a formal BH mass above
zero. Of those 31 UCDs, 19 also have their lower BH mass limits above
zero. That is, 19 UCDs ($\sim$40\% of all) formally require at the
1$\sigma$ level some additional dark mass to account for their
elevated dynamical M/L ratio. We re-iterate that our M/L threshold for
requiring dark mass assumes a 13 Gyr age for a canonical stellar
population. This is a conservative limit which may assign too low dark
mass fractions to some UCDs, if their light were dominated by a
stellar population substantially younger than 13 Gyr
(Sect.~\ref{singlepop} of this paper) or if they have been subject to
significant dynamical evolution (Kruijssen \&
Mieske~\citeyear{Kruijs10}).

Under the assumption that the IMF in UCDs is canonical (e.g. following
a Kroupa~\citeyear{Kroupa02} parametrisation), one can thus speculate
that about half of the UCD population may contain a central BH with a
mass sufficient to notably elevate the measured velocity
dispersion. In this context, the term 'notably' refers to the
observationally derived error bars of the ratio between final
dynamical M/L measurement and stellar population M/L, which includes a
general 0.3 dex error bar in [Fe/H]. The top left panel of
Fig.~\ref{BHrel_King} plots the estimated BH fraction in UCDs vs. the
1 $\sigma$ uncertainty of that estimate (lower limit). No BH mass
estimate is more accurate than $\sim$20\% relative error. Furthermore,
only eight of the UCDs in our sample have relative errors $<50\%$,
i.e. a $>2 \sigma$ significant BH mass estimate (seven of those are
high-mass UCDs with $M > 10^7 \msun$). This 2$\sigma$ significance
limit coincides with a BH mass fraction of about 10\%.

\subsubsection{BH occupation fraction in high-mass vs. low-mass UCDs}

\label{ocfrac}

It has been shown in Sect.~\ref{bimodal} that the
  distribution in $\Psi$ for high-mass UCDs is unimodal with a mean
  $\Psi \sim 1.7$, while for low-mass UCDs the distribution appears
  bimodal with peaks at $\sim$0.6 and $\sim$1.3. This is naturally
  also reflected in the BH mass distribution of the two
  sub-samples. Of the 19 UCDs that formally require a BH at the
  1$\sigma$ level, 13 are high-mass UCDs (2/3 of all high-mass UCDs)
  and 6 are low-mass UCDs (1/5 of low-mass UCDs). In contrast, within
  the 18 UCDs that have formal BH masses of 0, there is only one
  single high-mass UCD. See Table~\ref{tableall}. 

Under the assumption that all high-mass UCDs are naked nuclei our
data would thus suggest that 60-100\% of the UCD progenitor galaxies
in our sample possessed a SMBH.

Among the 30 low-mass UCDs, 13 have formal BH masses above zero and 17
UCDs have formal BH masses of zero. This approximate equipartition
between UCDs with and without BH is a consequence of the above
mentioned bimodal distribution in $\Psi$ for low-mass UCDs. As
mentioned in Sect.~\ref{bimodal}, this could be explained if the
transition mass regime $2 \times 10^6 < M < 10^7 \msun$ is populated
both by the high-mass extension of the canonical globular cluster
population, and the low-mass tail of tidally stripped nuclear
clusters. In the next subsection we elaborate more on this possibility
from a statistical point of view.



\subsection{Star clusters vs. tidally stripped nuclei}

In Mieske et al.~(\citeyear{Mieske12}) it was shown that the overall
number counts of UCDs are consistent with being a simple extrapolation
of the globular cluster luminosity function towards brighter
luminosities. An upper limit of $\sim$50\% of contamination from an
additional formation channel was derived in that work. 

Based on these numbers it is argued in Mieske et
al.~(\citeyear{Mieske12}) that it is {\it not necessary} to invoke an
additional formation channel for UCDs. We stress here that this is not
in contrast to the above formulated hypothesis that the most massive
UCDs originate (mainly) from tidally stripped dwarf galaxies, because
those massive UCDs are only a small fraction of the full UCD regime,
see next paragraph. Also, the contamination of up to 50\% of
  tidally stripped nuclei in the low-mass UCD regime $M < 10^7 \msun$
  as suggested in Sect.~\ref{distribution} and ~\ref{ocfrac} is still
  within the range 'allowed' by the upper contamination limit derived
  in Mieske et al.~(\citeyear{Mieske12}).

The respective mass threshold of $M \sim 10^7 \msun$ corresponds to a
luminosity of $M_V \simeq -11.5$ mag, when assuming M/L$\sim$3
(Fig.~\ref{massML}). This is about 1.3 mag brighter than the
luminosity limit of $M_V=-10.2$ mag adopted for UCDs in Mieske et
al.~(\citeyear{Mieske12}) - calculated for a mass of $M = 2 \times
10^6 \msun$ and a M/L=2 (Fig.~\ref{massML}). A typical globular
cluster luminosity function peaks around $M_V=-7.4$ mag and has a
width of $\sim$1.3 mag (Mieske et al.~\citeyear{Mieske12} and
references therein). Based on the normal distribution, the number of
objects with $M_V<-11.5$ mag (=3.15$\sigma$ away from the Gaussian
peak) is only 0.001 of the whole population, and only 6-7\% of the
entire UCD population with $M_V<-10.2$ mag (=2.15$\sigma$ away from
the peak). The fact that in Fig.~\ref{massML} the relative fraction of
massive UCDs with $M > 10^7 \msun$ is much larger ($\sim$40\%) is due
to the higher observational completeness in the bright luminosity
regime.

Fig.~5 of Mieske et al.~(\citeyear{Mieske12}) compares for the Fornax
cluster the cumulative distribution of galaxies in the expected UCD
progenitor mass range (Bekki et al.~\citeyear{Bekki03}) to the
cumulative distribution of present-day UCDs. The present-day number of
potential progenitor galaxies is more than an order of magnitude lower
than the total number of UCDs when considering all UCDs with
$M_V<-10.2$. The numbers of UCDs and potential progenitors become
comparable, and hence more consistent with theoretical predictions
(Henriques et al.~\citeyear{Henriq08},~\citeyear{Henriq10}), for UCD
luminosities $M_V \lesssim -11.5$ mag. This coincides with the limit
above which there appears systematic evidence for additional dark mass
in UCDs, and is again consistent with the notion that tidally stripped
galaxies dominate over the star cluster channel only in the high mass
UCD regime.

\subsection{The location of UCDs in the M$_{\rm BH}$-$L$ and M$_{\rm BH}$-$\sigma$ plane}

Further insight into the nature of UCDs can be gained by placing them
on the well known M$_{\rm BH}$-L and M$_{\rm BH}$-$\sigma$ relation
(e.g. Ferrarese \& Merritt~\citeyear{Ferrar00}), using the putative BH
masses estimated above. In Fig.~\ref{sigmass} we show in the left
panel the global velocity dispersions of the UCDs in our sample
vs. these BH masses, plotted in log-log-space. The right panel plots
the total luminosity vs. BH mass. For comparison we indicate central
black hole mass estimates in Galactic globular clusters from
L\"{u}tzgendorf et al. (\citeyear{Luetzg13}) with red
asterisks\footnote{ We note that the BH masses in GCs from the
  compilation in L\"{u}tzgendorf et al. (\citeyear{Luetzg13}) are
  generally less significant than 3$\sigma$, including some sources
  with only upper limits, as indicated in the figure. See also
  Anderson \& van der Marel~(\citeyear{Anders10}) and Lanzoni et
  al.~(\citeyear{Lanzon13}) who find significantly lower
  upper-mass-limits on the central BH in two Globular Clusters listed
  in L\"{u}tzgendorf et al. (\citeyear{Luetzg13}).}, and the galaxy
sample from McConnell et al. (\citeyear{McConn13}) as blue open
circles. We also show as dashed line a linear fit in log-log space to
the galaxy sample. The slope is 4.9 for the M$_{\rm BH}$-$\sigma$
relation, and 1.2 for the M$_{\rm BH}$-$L$ relation of the galaxy
sample.

In the M$_{\rm BH}$-$L$ plane UCDs are offset towards lower
luminosities by about a factor of 100 compared to the relation defined
by galaxies. In the M$_{\rm BH}$-$\sigma$ plane UCDs show an analogous
offset of about 2 dex in BH mass with respect to the relation defined
by galaxies. This suggests that in the framework of the tidal
stripping scenario, today's UCDs would have $\sim$1\% of the
luminosity of their potential progenitor galaxy. For nuclear clusters
in present-day galaxies, the right panel of Fig.~\ref{NCrelmass} shows
that such a ratio of $\sim$1\% is typical for nuclei that have BH mass
fractions of $\sim$10\%, as estimated for UCDs.

This lends further plausibility to the hypothesis that some UCDs are
created by tidal stripping, having kept the massive central black hole
as a relict tracer of their much more massive progenitor galaxies
(Bekki et al.~\citeyear{Bekki03}, Pfeffer \&
Baumgardt~\citeyear{Pfeffe13}). It is interesting to note in the right panel
of Fig.~\ref{sigmass} that the MW's nuclear cluster, if considered
as single object without its host galaxy, is consistent with the
postulated location of UCDs: the MW BH has about 1/10th of the nuclear
cluster's mass (Graham \& Spitler~\citeyear{Graham09} and references
therein).

A basic sanity check of this scenario is to combine the
factor 100 offset to the $M_{\rm BH}-$Luminosity relation of galaxies
with the 10-15\% BH mass fractions in UCDs. The implied BH mass
fraction in UCD progenitor galaxies would then be of order
$\sim$0.1-0.2\%, indeed close to the 0.2\% BH mass fractions found
generally in galaxies (e.g. Ferrarese et
al.~\citeyear{Ferrar06}). This is, however, not an independent
argument per se, since the location of the $M_{\rm BH}-$Luminosity
relation for galaxies is what defines the overall $\sim$0.2\% BH mass
fraction.

\begin{figure}[h!]
\begin{center}
\includegraphics[width=4.3cm]{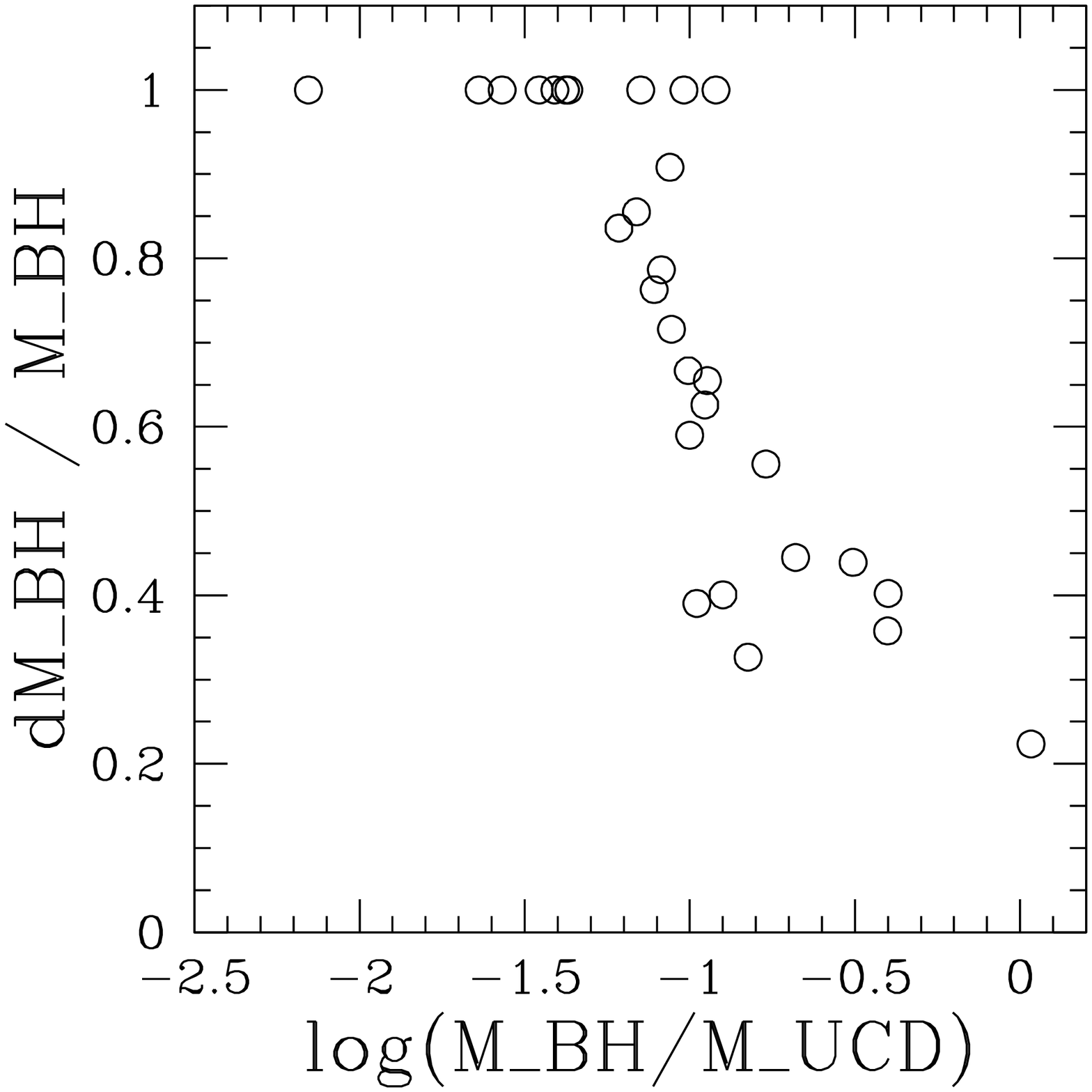}
\includegraphics[width=4.3cm]{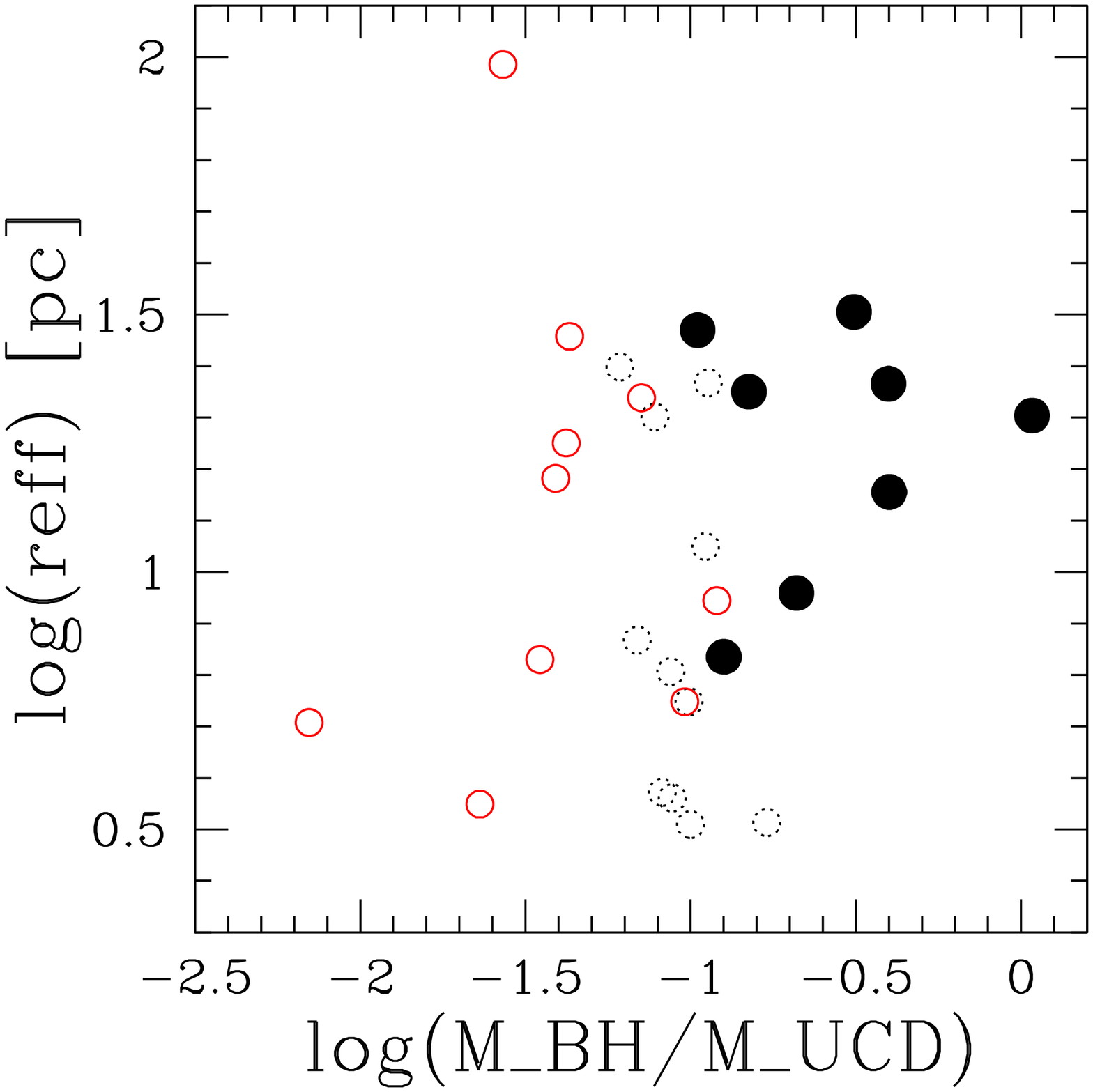}\\
\includegraphics[width=4.3cm]{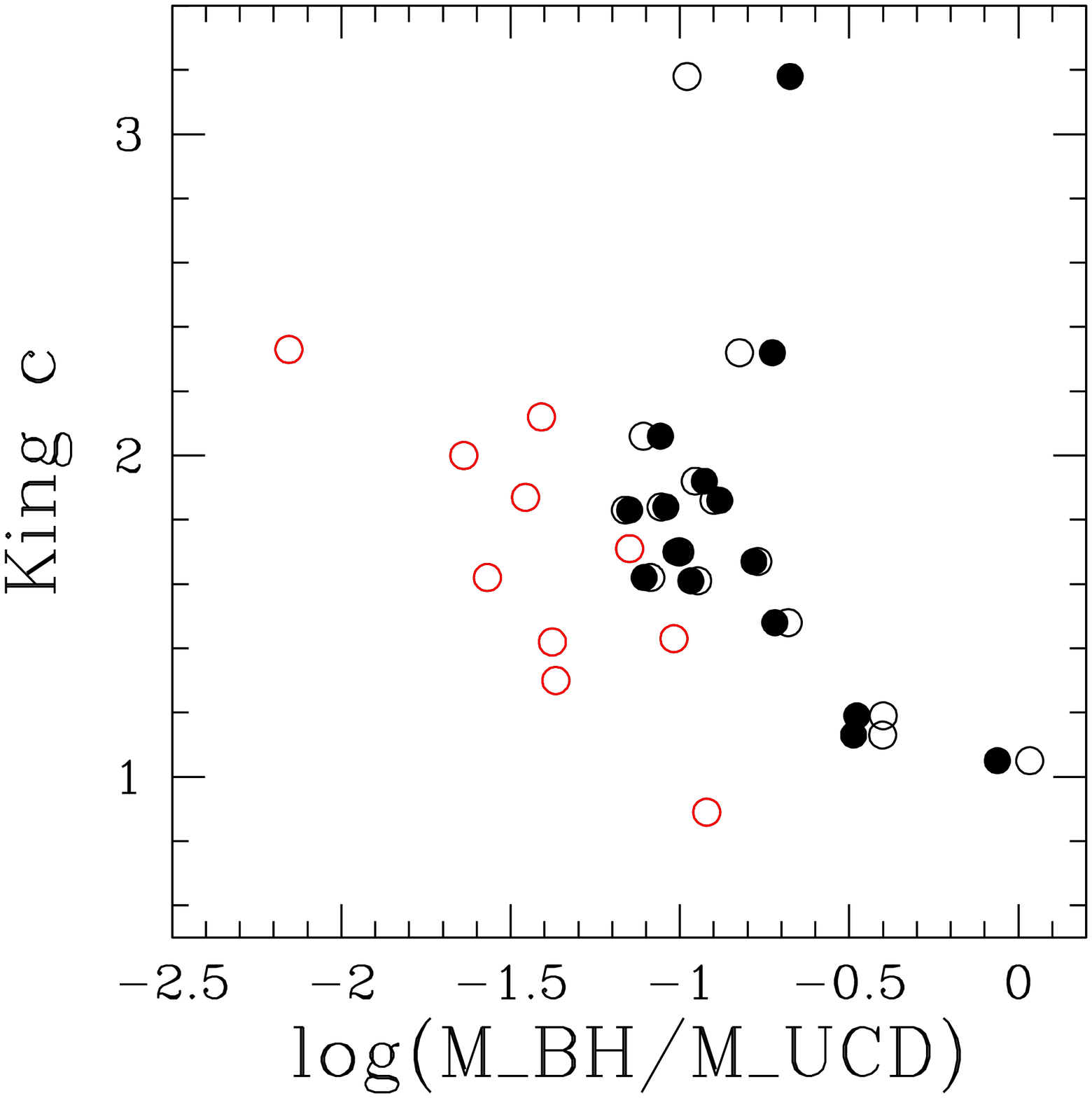}
\includegraphics[width=4.3cm]{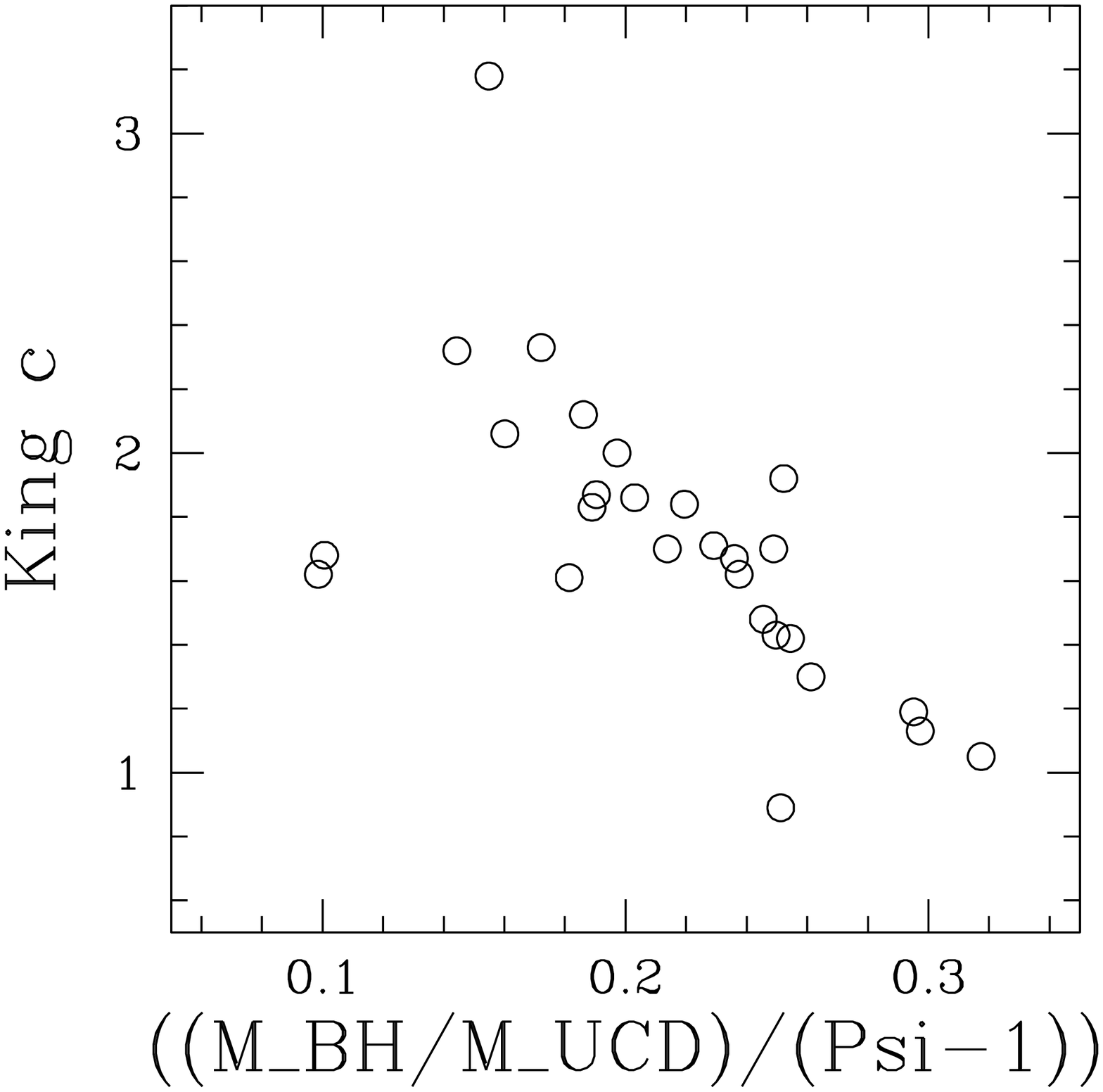}
  \caption{{\bf Top left panel}: BH mass fraction of UCDs plotted
    vs. its relative error. Sources with relative error of one are
    those where the lower error bar of the BH mass is consistent with
    0. BH mass estimates more significant than $\sim$2$\sigma$ require
    $\gtrsim$10\% BH mass. {\bf Top right panel:} BH mass fraction
    plotted vs. the effective radius in arcseconds. Red open circles
    are those objects where the lower error bounds of the dynamical
    mass measurements are consistent with zero BH mass. Dotted black
    circles denote those UCDs whose BH mass estimate is different from
    zero between 1 and 2$\sigma$ significance. Filled black circles
    indicate those UCDs with black hole masses more than 2$\sigma$
    significant. These are those with relative error $<0.5$ in the
    top left panel. {\bf Bottom left panel}: BH mass fraction in
    UCDs is plotted vs. the concentration parameter $c$ of their
    surface brightness distribution, when parametrised as a King
    profile. The filled dots indicate the BH mass fraction assuming
    that all UCDs have the same King parameter $c$ of 1.75, taking
    into account the correlation between $c$ and BH mass as shown in
    the right panel. {\bf Bottom right panel}: The y-axis shows the
    concentration parameter $c$ as in the left panel. The x-axis shows
    the ratio between the BH mass fraction and the additional
    dynamical mass derived under the assumption that mass follows
    light, ($\Psi -1$). The fact that this ratio is generally below 1 is because
    the putative BH is located in the center of the UCD, as opposed to
    a uniformally distributed additional mass component. The overall
    trend is due to the fact that for more centrally concentrated
    objects, the effect of the BH on the integrated velocity
    dispersion increases. On average, a central BH of a given mass has the
    same effect on the global velocity dispersion as 4-5 times that mass
    distributed uniformally.}
\label{BHrel_King}
\end{center}
\end{figure}

\subsection{Dependence of UCD BH mass on structural parameters}

It is commonly argued that the presence of a central black hole in the
center of a galaxy can affect the stellar distribution, up to the
point where a previously existing nucleus is entirely evaporated
(e.g. C\^{o}t\'{e} et al.~\citeyear{Cote06}, Bekki \& Graham~\citeyear{Bekki10},
Antonini et al.~\citeyear{Antoni13}). It is thus interesting to test
whether structural parameters of the UCDs in our sample correlate with
the hypothetical BH mass fraction.  In Fig.~\ref{BHrel_King} we show
two plots that relate the hypothetical BH mass in UCDs to the UCDs'
structural parameters.

The top right panel plots the relative BH mass against the effective
radius. No dependence between both entities is found.

The bottom left panel of Fig.~\ref{BHrel_King} depicts the relative BH
mass as a function of central concentration of the light profile
(parametrised by the King concentration c). Considering those sources
whose hypothetical BH mass is significant beyond the 1$\sigma$
individual error bar, a trend may be seen. A least
squares fit to the data points reveals a correlation with a slope
different from zero at the 2.8$\sigma$ level significance level. The
Spearman correlation coefficient is $\rho=-0.543$, with a probability
of 3\% that there is no correlation. Also a Pearson and a Kendall test
yield similar probabilities of 2-3\%. The formal significance of this
trend is thus between 2-3$\sigma$.

The bottom right panel depicts the King concentration parameter
against the ratio between BH mass fraction and dynamical M/L surplus,
$\Psi -1$. This ratio illustrates by how much more a {\it central}
BH of a given mass influences the global velocity dispersion compared
to a {\it uniformly} distributed dark mass component. The effect of a
central BH on the global velocity dispersion is typically a factor of
4-5 higher than the effect of the same amount of mass in uniform
distribution. Aside this general offset, there is also a clear trend
in the sense that a central BH alters the global velocity dispersion
stronger for sources with higher King concentration parameter. This is
qualitatively expected, since a higher central concentration implies a
larger relative number of stars within the sphere of influence of the
BH.

The trend shown in the bottom right panel may also, at least
partially, explain the trend between King concentration parameter and
BH fraction seen in the bottom left panel. If one were to assume,
conservatively, that all UCDs have the same intrinsic concentration
$\sim$ 1.75, and that the distribution of concentrations measured in
the literature are merely due to measurement uncertainties, then the
resulting BH mass fractions would need to be renormalised to that same
average concentration, based on the relation seen in the bottom right
panel. The result of such an exercise is shown by the
small filled circles in the bottom left plot: the BH mass fraction are
'corrected' to an assumed true King concentration of 1.75. For this
case, the correlation between King concentration and BH mass fraction
is removed (the formal significance drops to the 1.4\% level). We note
that typically quoted uncertainties of King concentration (which
itself is a logarithm) are 0.1 to 0.2 in log space, while the overall
sample shows a King concentration scatter of 0.5 dex. The above
exercise thus illustrates an extreme assumption, but goes in the
direction as to refrain from a final conclusion on the significance of
a possible trend between concentration and BH mass fraction.

\begin{figure*}
\includegraphics[width=8.6cm]{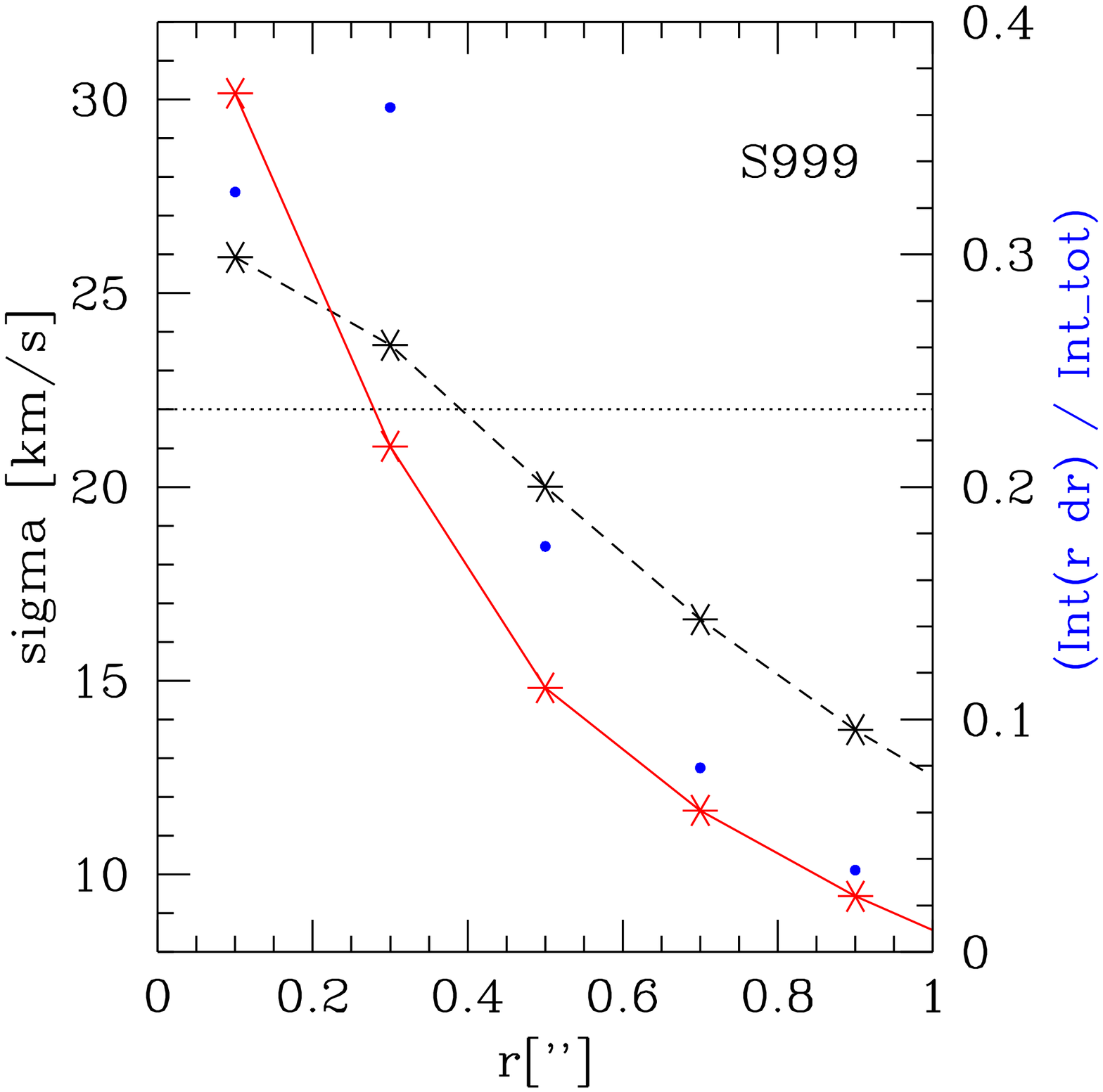}
\includegraphics[width=8.6cm]{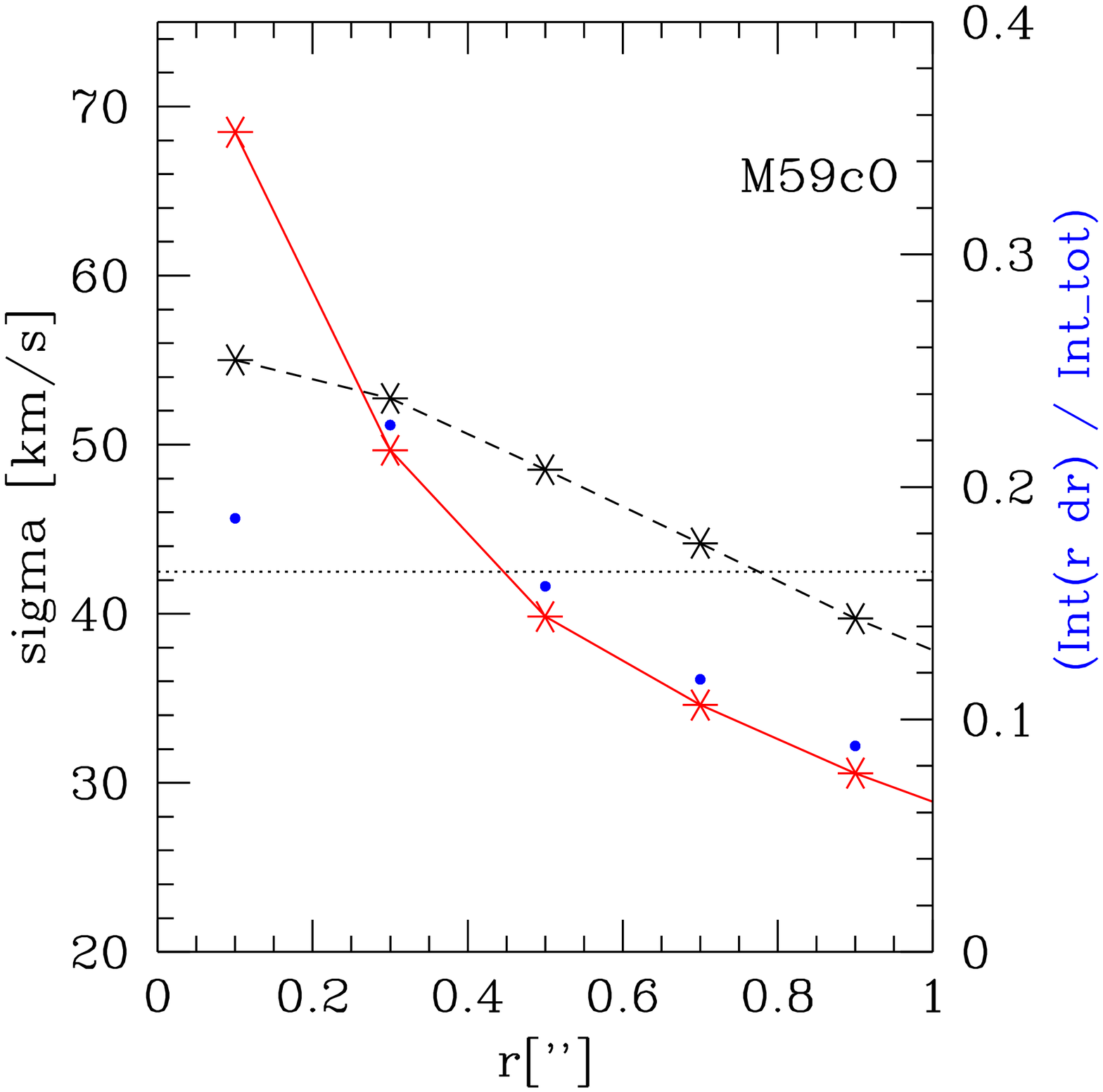}
\caption{Expected velocity dispersion profile for the two UCDs with
  the highest predicted BH masses. {\bf Left panel:} Virgo UCD S999
  (Ha\c{s}egan et al.~\citeyear{Hasega05}), for which a central BH of
  $\sim 2.5 \times$ 10$^7 \msun$ is predicted. {\bf Right panel:}
  Virgo UCD M59cO (Chilingarian \& Mamon ~\citeyear{Chilin08}), for
  which a central BH of $\sim 5 \times$ 10$^7 \msun$ is predicted. In
  both plots, the assumed spatial resolution is 0.2$''$, corresponding
  to the typical resolution possible in ground based AO assisted
  spectroscopy. The measured global velocity dispersion is indicated
  by the horizontal dotted line. The dashed (black) line is the
  velocity dispersion profile that reproduces the observed global
  velocity dispersion for the assumption of a {\it uniformally
    distributed} dark mass component on top of the photometric
  mass. The solid (red) line is the velocity dispersion profile that
  reproduces the observed global velocity dispersion for the
  assumption of a {\it central point mass} on top of the photometric
  mass. The small (blue) dots indicate the fraction of light from the
  source that is contained within each ring around the indicated
  values of r, with the right y-axis showing the scale.}
\label{profile}
\end{figure*}

\section{Conclusions and Outlook}
\label{summary}

Massive UCDs ($\M>10^7\msun$) have elevated dynamical M/L ratios
compared to expectations from stellar population models, with an
average ratio $\Psi=(M/L)_{\rm dyn}/(M/L)_{\rm pop} = 1.7\pm
0.2$. This implies notable amounts of dark mass in them. 

We find that, on average, central BH masses of 10-15\% of the UCD mass
can explain these elevated dynamical M/L ratios of UCDs, with individual
BH masses between 10$^5 \msun$ and several 10$^7 \msun$
(Fig.~\ref{BHrelmass}). In the $M_{\rm BH}-$Luminosity plane, UCDs are
offset by a factor of $\sim$100 in BH mass from the relation defined
by galaxies (Fig.~\ref{sigmass}).

These findings fit well to the scenario that massive UCDs are tidally stripped
nuclei of originally much more massive galaxies, and their putative
central black holes are a fossil relict of their progenitor:

\begin{itemize}

\item The estimated BH mass fractions of 10-15\% agree with those
  found in present-day nuclei of galaxy bulges (Fig.~\ref{NCrelmass}) that have
  the masses expected for UCD progenitor galaxies (M$\sim$$10^9 \msun$;
  Bekki et al.~\citeyear{Bekki03}, Pfeffer \&
  Baumgardt~\citeyear{Pfeffe13}).

\item The factor 100 offset to the $M_{\rm BH}-$Luminosity relation
  would suggest UCD progenitor masses M$\sim$$10^9
  \msun$, again consistent with expectations within the tidal
  stripping scenario.
\end{itemize}

The elevated dynamical M/L ratios of UCDs are most prominent for UCD
masses above $\sim$$10^7 \msun$, which corresponds to $\lesssim 10\%$
of the overall UCD population. This relatively small fraction is
consistent with the finding that the overall number of UCDs matches
well the extrapolated globular cluster luminosity function (Mieske et
al.~\citeyear{Mieske12}). {We also find tentative evidence for a
  bimodal $\Psi$ distribution in low-mass UCDs ($2 \times
  10^6<\M<10^7\msun$), suggestive of an overlap of globular clusters
  and tidally stripped nuclei in this regime. 

From these findings a picture of two UCD formation channels emerges: a
'globular cluster channel' important mainly for UCDs with $M \lesssim
10^7 \msun$; and, tidal transformation of massive progenitor galaxies
which dominates for UCDs $M \gtrsim 10^7 \msun$ and still contributes
for lower UCD masses. 

We re-iterate that a massive BH as relict tracer of a massive
progenitor is of course only one possibility to explain the elevated
M/L. Other scenarios include a globally different IMF (Mieske \&
Kroupa \citeyear{Mieske08b}; Dabringhausen et
al.~\citeyear{Dabrin12}), or highly concentrated dark matter
(Ha\c{s}egan et al.~\citeyear{Hasega05}, Goerdt et
al.~\citeyear{Goerdt08}, Baumgardt \& Mieske~\citeyear{Baumga08}).

The next step in assessing the plaubsility of the BH scenario are more
direct observational tests. One clear possibility is to measure 
spatially resolved velocity dispersion profiles. As noted in the
Introduction, a first such measurement was presented for the Fornax
UCD3 by Frank et al.~(\citeyear{Frank11}), based on IFU observations at
very good natural seeing (0.5-0.6$"$ resolution). In that
investigation, no indication for the characteristic expected rise of
the velocity dispersion profile at small radii was detected. Due to
its large intrinsic size of $\sim$1$"$, UCD3 is one of the very few
UCDs suited for such natural seeing tests. However, at the same time
it is one of the UCDs whose dynamical M/L are fully consistent with a
canonical stellar population without a BH (Table~\ref{tableall}).

Therefore, observational efforts should now focus on adaptive optics
assisted spectroscopy with higher spatial resolution, to be able to
target also those UCDs with smaller intrinsic sizes $\lesssim
0.5"$. In Fig.~\ref{profile} we show the predicted velocity dispersion
profiles for the two UCDs with highest predicted BH masses, S999 and
M59cO in Virgo, assuming a spatial resolution of 0.2$"$. From the
plots one can distill the obervational requirement to distinguish
between the profile with a central BH of the predicted mass and a {\it
  mass-follows-light} case. Three independent radial measurements out
to $r \sim 0.5"$ would require a precision $\Delta \sigma \sim$ 5 km/s
at $\sigma=$20 km/s for S999, and $\Delta \sigma \sim$ 10 km/s at
$\sigma=$50 km/s for M59cO. Especially for M59cO this is feasible
with current instrumentation, e.g. LGS IFU spectropscopy using
NIFS@GEMINI or SINFONI@VLT. In Fig.~\ref{profile} we also indicate the
fraction of the total flux which is contained in each ring around a
given r, based on the literature surface brightness profile and an
assumed spatial resolution of 0.2$"$ achievable with adaptive
optics. For the innermost three radial bins up to $r\sim 0.5"$, the
flux per bin is relatively constant, and only drops for larger
radii. A direct observational detection of BH signatures thus seems
possible for the most emblematic UCDs.

Another possibility for BH detection in UCDs is to analyse their X-ray
and Radio emission, looking for the signatures typical for the
accretion onto BHs (e.g. Gallo et al.~\citeyear{Gallo03}). Several
recent studies have reported evidence for stellar mass or intermediate
mass black holes in globular clusters and dwarf galaxies from such
analyses (e.g. Maccarone et al.~\citeyear{Maccar07}, Farrell et
al.~\citeyear{Farrel09}, Strader et
al.~\citeyear{Strade12}\&~\citeyear{Strade13}; but see also Maccarone
et al.~\citeyear{Maccar08}), which makes a systematic survey in the
UCD regime an interesting prospect.

\vspace{0.2cm}
\noindent We conclude that central BHs as relict tracers of massive
progenitors are a plausible explanation for the elevated dynamical M/L
ratios of massive UCDs (M$\gtrsim 10^7 \msun$).

\begin{acknowledgements} M.~J.~F. gratefully acknowledges support from the
German Research Foundation (DFG) via Emmy Noether Grant Ko 4161/1, and
thanks ESO for support from the Director General Discretionary
Fund. H.B. acknowledges support from the Australian Research Council
through Future Fellowship grant FT0991052.

\end{acknowledgements}

\bibliographystyle{aa}
\bibliography{paperucdbh_accepted}

\end{document}